\documentclass[10pt,twocolumn,aps,english,pra,superscriptaddress,longbibliography]{revtex4-2} 

\pdfoutput=1
\usepackage[utf8]{inputenc}
\usepackage[english]{babel}
\usepackage[T1]{fontenc}
\usepackage{amsmath}
\usepackage{hyperref}
\usepackage{physics}
\usepackage{tikz}
\usepackage{lipsum}
\usepackage{bm}
\usepackage{bbm}
\usepackage{mathrsfs}
\usepackage{amsfonts}
\usepackage[T1]{fontenc}
\usepackage{amsmath}
\usepackage{amssymb}
\usepackage{graphicx}
\usepackage{babel}
\usepackage{booktabs}
\usepackage{comment}
\usepackage{tabu}
\usepackage{subfigure}
\usepackage{physics}
\usepackage{extarrows}
\usepackage{xcolor}
\usepackage{placeins}
\usepackage{bm}

\usepackage{xcolor}
\newcommand{\yk}[1]{#1}

\begin{document}

\title{Astronomical interferometry using continuous variable quantum teleportation }

\author{Yunkai Wang} 
\altaffiliation{Current affiliation: Perimeter Institute for Theoretical Physics, Waterloo, Ontario N2L 2Y5, Canada; Institute for Quantum Computing and Department of Applied Mathematics, University of Waterloo, Waterloo, ON N2L 3G1, Canada}
\affiliation{IQUIST, University of Illinois at Urbana-Champaign, Urbana, IL 61801 USA}
\affiliation{Department of Physics, University of Illinois at Urbana-Champaign, Urbana, IL 61801 USA}

\author{Yujie Zhang} 
\altaffiliation{Current affiliation: Perimeter Institute for Theoretical Physics, Waterloo, Ontario N2L 2Y5, Canada; Institute for Quantum Computing and Department of Physics \& Astronomy, University of Waterloo, Waterloo, ON N2L 3G1, Canada}
\affiliation{IQUIST, University of Illinois at Urbana-Champaign, Urbana, IL 61801 USA}
\affiliation{Department of Physics, University of Illinois at Urbana-Champaign, Urbana, IL 61801 USA}

\author{Virginia O.~Lorenz} 
\affiliation{IQUIST, University of Illinois at Urbana-Champaign, Urbana, IL 61801 USA}
\affiliation{Department of Physics, University of Illinois at Urbana-Champaign, Urbana, IL 61801 USA}

\begin{abstract}

We propose a method to build an astronomical interferometer using continuous variable quantum teleportation to overcome the transmission loss between distant telescopes. The scheme relies on two-mode squeezed states shared by distant telescopes as entanglement resources, which are distributed using continuous variable quantum repeaters. We find the optimal measurement on the teleported states, which uses beam-splitters and photon-number-resolved detection. Compared to prior proposals relying on discrete states, our scheme has the advantages of using linear optics to implement the scheme without wasting stellar photons, and making use of multiphoton events, which are regarded as noise in previous discrete schemes. \yk{We also outline the parameter regimes in which our scheme outperforms the direct detection method, schemes utilizing distributed discrete-variable entangled states, and local heterodyne techniques.}
\end{abstract}

\maketitle

\section{Introduction}

Interferometric imaging is a widely used method in astronomy that employs multiple spatially separated telescopes for enhanced resolution and sensitivity. \yk{By synthesizing data from these telescopes, the method effectively attains the resolution of a much larger aperture, with the maximum resolution determined by the longest baseline between the telescopes.} It is based on the Van Cittert-Zernike theorem \cite{zernike1938concept}, which shows one can determine the Fourier component of the intensity distribution in the source plane by interfering with the light received at different locations in the image plane and measuring the mutual coherence function. The successful application of interferometer arrays in astronomy, especially at radio frequencies, provides a very powerful tool for high-resolution imaging. For example, the first image of a supermassive black hole at the center of the Messier 87 Galaxy was provided by a radio interferometer array \cite{collaboration2019first}. However, unlike radio wavelengths, for which the light is directly measured and recorded separately at each telescope \cite{wilson2009tools}, at optical wavelengths the light signals received at different locations in the interferometer array are at frequencies too high to be recorded, so instead they are brought together to interfere in what is referred to as the direct-detection method \cite{monnier2003optical}. However, in this method there is unavoidable transmission loss while bringing light from distant telescopes together, and this limits the length of the baseline and hence the resolution of optical interferometer arrays. It is also possible to measure the coherence function through local measurements with a shared phase reference, but for weak thermal sources at optical wavelengths the mean photon number per temporal mode is much less than one, and it is not possible to distinguish the vacuum and single photon states locally. This strongly degrades the sensitivity of local measurements at each telescope \cite{tsang2011quantum}.

Several proposals have been made recently that overcome transmission loss based on quantum networks \cite{gottesman2012longer,khabiboulline2019optical,khabiboulline2019quantum, Huang2022, Czupryniak2023}. Taking advantage of entanglement resources provided by the quantum network, the coherence function can be measured without directly interfering the light and with sensitivity comparable to the direct-detection method. These proposals can be understood qualitatively as leveraging teleportation from one telescope to another using entanglement resources shared by the distant telescopes. Transmission loss will affect the distribution of the entanglement resource, which can be overcome by quantum repeaters. A quantum repeater distills high-fidelity entangled states from many copies of distributed noisy entangled states between nearby quantum nodes and creates long-distance entanglement from short-distance entanglement using entanglement swapping \cite{sangouard2011quantum}. Although this in principle means we can arbitrarily increase the baseline between telescopes, these proposals have their own difficulties in implementation. Reference \cite{gottesman2012longer} requires an excessive amount of distributed entangled photons.  Reference \cite{khabiboulline2019optical,khabiboulline2019quantum, Huang2022, Czupryniak2023} exploits quantum memory to encode the arrival time of the stellar photons to avoid the wasting of entanglement resources when a vacuum state is received. This approach introduces the extra difficulties of implementing two-qubit quantum gates and requires reliable quantum memories. By contrast, we are mainly focused on a new entanglement-assisted protocol that does not require memories at the telescopes and is compatible with a near-term quantum network. Along this direction, protocols using $N$-copy single-photon entangled states and other entangled states have been developed \cite{Czupryniak2022, Marchese2023, zhang2025, modak2024}, demonstrating a significant quantum enhancement over the original quantum telescope scheme \cite{gottesman2012longer}, particularly in scenarios with transmission loss \cite{Marchese2023}. However, these phase-estimation protocols based on $N$-copy entangled states have only been applied to sensing astronomical objects approximated as point sources, and only apply when the astronomical source can be approximated as a weak thermal source. In this letter, we introduce a novel continuous-variable entanglement-assisted protocol designed to achieve enhanced performance in quantum telescopy across a broader range of sources, varying both in intensity and intensity distribution.

In this paper, we extend the concept of quantum-network-based astronomical interferometry to another version of quantum teleportation, namely, continuous variable (CV) quantum teleportation \cite{vaidman1994teleportation,braunstein1998teleportation,pirandola2006quantum}. Compared to Ref.  \cite{gottesman2012longer,khabiboulline2019optical,khabiboulline2019quantum}, where the weak thermal light received from the astronomical source is approximated as discrete states, namely, vacuum or single photon states, here we directly work with the exact thermal state, which is a Gaussian state with representation in terms of Gaussian functions \cite{weedbrook2012gaussian}. Our scheme relies on the analog of Einstein-Podolsky-Rosen (EPR) states in continuous variables, i.e. the two-mode squeezed states. Here we discuss the sensitivity of our scheme under transmission loss and construct optimal measurements for estimating the coherence function from the teleported state.  \yk{We find that the required repetition rate to cover all temporal modes is approximately 150 GHz at wavelength $\lambda=800\;\text{nm}$, much higher than state-of-the-art pulsed squeezing with 163 MHz repetition rate and  6.8 dB squeezing level \cite{dong2008experimental}.}
Nevertheless, our method still provides a meaningful alternative for building an astronomical interferometer that may be more feasible than other quantum network protocols, depending on the development of quantum repeaters. In particular, our scheme can exploit multiphoton events that are discarded as noise in Ref.~\cite{gottesman2012longer,khabiboulline2019optical,khabiboulline2019quantum}, which can provide an advantage when imaging a stronger astronomical source or at a longer wavelength. \yk{Assuming ideal implementation of each scheme, we explicitly compare our CV quantum network-based scheme with the original proposal based on discrete-variable (DV) quantum networks \cite{gottesman2012longer}, and local heterodyne detection. Interestingly, we identify a regime corresponding to the source having intermediate strength, in which our approach outperforms both of these schemes. Furthermore, we account for imperfections in the practical implementation of our CV quantum network-based scheme and compare its performance with direct detection, the DV quantum network-based scheme \cite{gottesman2012longer}, and local heterodyne detection under realistic conditions. We show that even in the non-ideal case, our scheme can still provide advantages in certain scenarios.}

\section{Teleportation of stellar light}

\begin{figure}[!tb]
\begin{center}
\includegraphics[width=1\columnwidth]{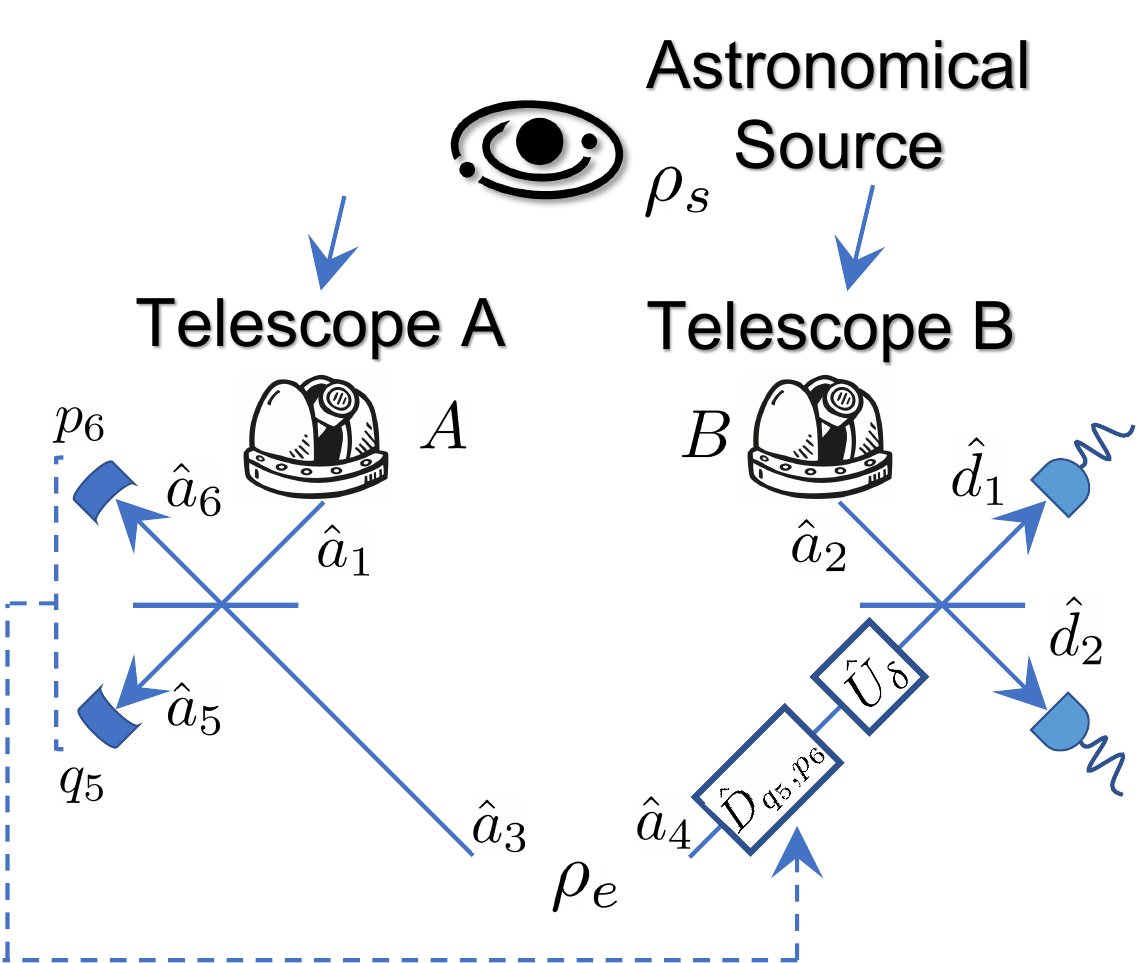}
\caption{Setup for astronomical interferometry using continuous variable quantum teleportation. $\rho_s$ is the stellar state from the astronomical source and $\rho_e$ is the two-mode squeezed state. The dashed line indicates classical communication from telescope A to telescope B about the outcome $q_5$, $p_6$ of homodyne detection. $\hat{D}_{q_5,p_6}$ is the displacement conditional on $q_5$, $p_6$. $\hat{U}_\delta$ is the phase delay added for the measurement of the teleported state. }
\label{set_up}
\end{center}
\end{figure}

In this section, we consider the CV teleportation of stellar light from one telescope to another telescope. As shown in Fig.~\ref{set_up}, we assume the bipartite thermal state $\rho_s$ of mode $\hat{a}_{1,2}$ from the astronomical source is received by two telescopes $A,B$ in an interferometer array \cite{mandel1995optical}, with the form:
\begin{equation}\begin{aligned}
\label{rho}
\rho_s=\int &\frac{d^2\alpha d^2\beta}{\pi^2\det \Gamma} \ket{\alpha,\beta}\bra{\alpha,\beta}\\
&\times\exp[-\begin{pmatrix}
\alpha* & \beta^*
\end{pmatrix}\Gamma^{-1}\begin{pmatrix}
\alpha \\ 
\beta
\end{pmatrix}]    
\end{aligned}
\end{equation}
\begin{equation}
\Gamma=\frac{\epsilon}{2}\begin{pmatrix}
1 & g\\
g^* & 1
\end{pmatrix},
\end{equation}
where $g=|g|e^{i\theta}$ is the coherence function we want to measure and $\epsilon$ is the mean photon number per temporal mode. Any  Gaussian state can be completely described with its mean value $x_i=\tr(\rho_s\hat{x}_i)$ and covariance matrix $V_{ij}=\frac{1}{2}\tr [\rho_s(\Delta \hat{x}_i\Delta \hat{x}_j+\Delta \hat{x}_j\Delta \hat{x}_i)]$, where $\hat{\vec{x}}=(\hat{q}_1,\hat{p}_1,\hat{q}_2,\hat{p}_2)^T$, $\hat{q}_i=(\hat{a}_i+\hat{a}_i^\dagger)/\sqrt{2}$, $\hat{p}_i=(\hat{a}_i-\hat{a}_i^\dagger)/\sqrt{2}i$, $\Delta \hat{x}_i=\hat{x}_i-x_i$ \cite{weedbrook2012gaussian}. The mean value $x_i$ and covariance matrix $V_{ij}$ for $\rho_s$ are
\begin{equation}
\begin{aligned}\label{V}
&x_i=0\\
&V=\frac{1}{2}\left[\begin{matrix}
1+\epsilon & 0 & \epsilon |g|\cos\theta & -\epsilon |g|\sin\theta\\
0 & 1+\epsilon & \epsilon|g|\sin\theta & \epsilon |g|\cos\theta\\
\epsilon |g|\cos\theta & \epsilon|g|\sin\theta & 1+\epsilon & 0\\
-\epsilon|g|\sin\theta & \epsilon |g|\cos\theta & 0 & 1+\epsilon
\end{matrix}.
\right]
\end{aligned}    
\end{equation}

\yk{In order to teleport the state received at telescope A to telescope B, we send the two-mode squeezed state $\rho_e = \ket{\text{TMSV}}\bra{\text{TMSV}}$, where $\ket{\text{TMSV}} = \hat{S}_2(r) \ket{0,0}$, with the squeezing operator given by $\hat{S}_2(r) = \exp\left( r (\hat{a}_3^\dagger \hat{a}_4^\dagger - \hat{a}_3 \hat{a}_4) \right)$, where $r$ is the squeezing parameter, from the entanglement source to the two telescopes through two lossy channels with the same transmission coefficient $T$. }
For the case where we need a quantum repeater, we just need to consider the output of the repeater with the effective transmission coefficient $T$ and use the same results derived below.

We follow the standard CV quantum teleportation scheme discussed in \cite{vaidman1994teleportation,braunstein1998teleportation,pirandola2006quantum}. As shown in Fig.~\ref{set_up}, we first combine the mode $\hat{a}_1$ of the stellar state and the mode $\hat{a}_3$ of the two-mode squeezed state on a beam-splitter at telescope A. We then perform homodyne detection on the two output ports of the beam-splitter and obtain the outcome $q_5$, $p_6$. The outcome $q_5$, $p_6$ is communicated to telescope B using a classical channel. We then implement a correction by displacing the state on telescope B.
The  teleported state can be described by its Wigner function as \cite{duan1997influence,chizhov2001propagation,scheel2001entanglement,chizhov2002continuous,ban2002continuous}
\begin{equation}\begin{aligned}
W(&\alpha_2,\alpha_4)=\frac{1}{2\pi+2\pi (e^{-2r}-1)T^2}\int d^2\alpha W(\alpha,\alpha_2)\\
&\times \exp\{-\frac{1}{2+2(e^{-2r}-1)T^2}[(p-p_4)^2+(q-q_4)^2]\}.
\end{aligned}\end{equation}
which is equivalent to the state after a thermalizing channel, as derived in Ref.~\cite{ban2002continuous}.
Its mean value and covariance matrix are
\begin{equation}
\begin{aligned}
&x_i=0,\\
&V'=V+2e^{-2r'}\left[\begin{matrix}
1 & 0 & 0 & 0\\
0 & 1  & 0 & 0\\
0 & 0 & 0 & 0 \\
0 & 0 & 0 & 0
\end{matrix}\right],
\end{aligned}    
\end{equation}
where $V$ is the covariance matrix of stellar light in Eq.~\ref{V}, and we have defined \yk{the effective squeezing parameter $r'$ such that $e^{-2r'}=1-T^2+e^{-2r}T^2$ for convenience.} Since we have $r=r'$ as $T\rightarrow1$, $r'$ can be regarded as the effective squeezing level after introducing loss to the channel. Notice that the CV teleportation only introduces some noise $2e^{-2r'}$ in the diagonal elements of the covariance matrix.  As $r'\rightarrow \infty$,  the quantum teleportation is ideal, i.e. $V'=V$. To improve the effective squeezing parameter $r'$, we need to increase both the transmission coefficient $T$ and the original squeezing parameter $r$. In particular, $1-T^2$ has to be comparable to $e^{-2r}$, otherwise increasing $r$ will not have a significant improvement on the effective squeezing parameter $r'$.

\section{Measurement on the teleported states}

\begin{figure}[!tbh]
\begin{center}
\includegraphics[width=1\columnwidth]{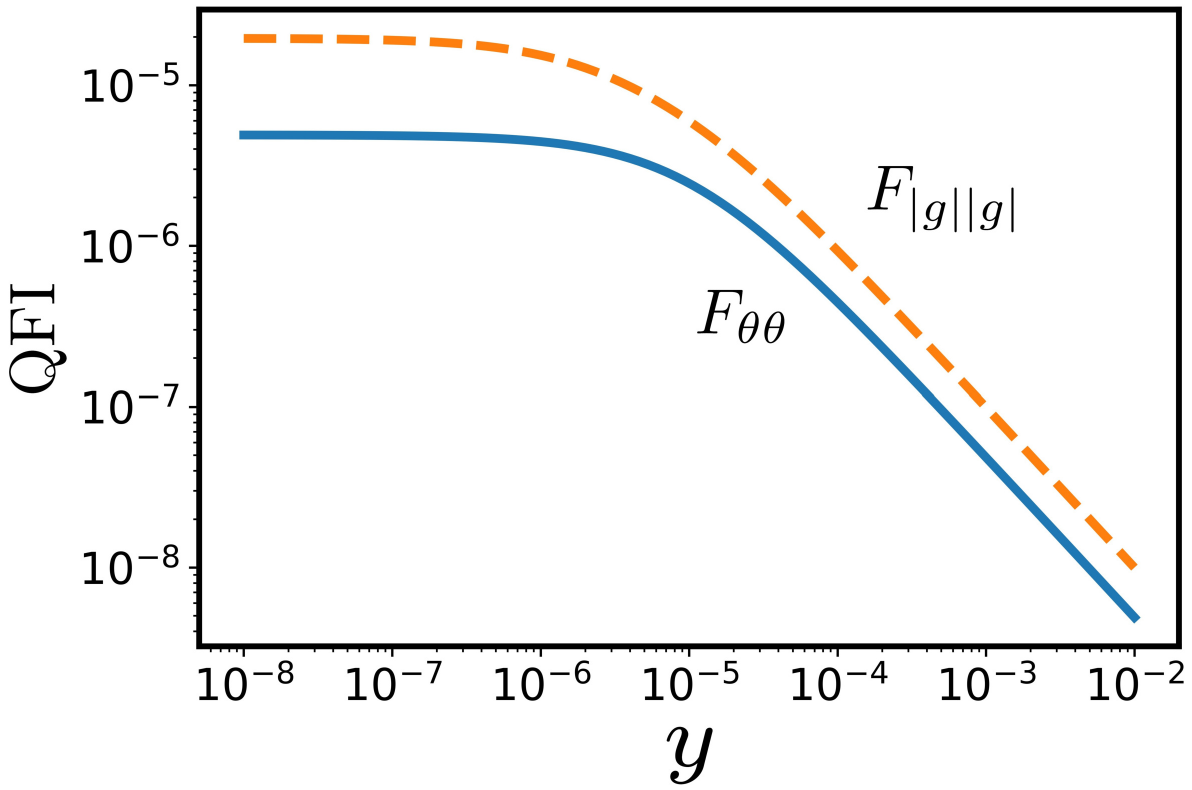}
\caption{The quantum Fisher information (QFI) for estimating the phase $\theta$ and amplitude $|g|$ of the coherence function as a function of the squeezing level $y$. The parameters are chosen as $|g|=0.7$, $\epsilon=10^{-5}$.}
\label{qfi}
\end{center}
\end{figure}

We now consider how well can we estimate the coherence function $g$ by measuring the nonideal teleported state. For the estimation of a set of parameters $\bm{z}=(\theta,|g|)^T$, the quantum Cram\'{e}r-Rao bound (QCRB) \cite{helstrom1976quantum} of estimating $\theta$ and $|g|$ is given by the quantum Fisher information (QFI) $F$: ${M_{\bm{z}}\geq F^{-1}}$, with its $(\mu,\nu)$ element ${[M_{\bm{z}}]_{\mu \nu}=\mathbb{E}\left[(\bm{z}_\mu-\check{\bm{z}}_\mu)(\bm{z}_\nu-\check{\bm{z}}_\nu)\right]}$, where $\check{\bm{z}}_\mu$ is the unbiased estimator of the $\mu$-th unknown parameter. The QCRB is a fundamental bound of the sensitivity optimized over all possible measurements and estimators. We here calculate the QFI to quantify the performance of our scheme. For a Gaussian state, the QFI $F_{ij}$ can be derived from its mean value $x_i$ and covariance matrix $\Sigma$ ( $\Sigma=\frac{1}{2}\tr[\rho(\mathbf{a}_i\mathbf{a}_j+\mathbf{a}_j\mathbf{a}_i)]$ is an equivalent form of covariance matrix $V$, where $\mathbf{a}=[a_1,a_1^\dagger,a_2,a_2^\dagger]$) \cite{monras2013phase,gao2014bounds}. Note that $\Sigma$ and $V$ are defined in terms of $\hat{a}_i$ and $\hat{x}_i$, respectively, and can be readily converted between each other. This leads to
\begin{equation}\label{QFI_formula}
F_{i j}=\frac{1}{2} \mathfrak{M}_{\alpha \beta, \mu \nu}^{-1} \partial_{j} \Sigma_{\alpha \beta} \partial_{i} \Sigma_{\mu \nu}+\Sigma_{\mu \nu}^{-1} \partial_{j} x_{\mu} \partial_{i} x_{\nu},
\end{equation}
where ${\mathfrak{M}=\Sigma \otimes \Sigma+\frac{1}{4} \Omega \otimes \Omega}$, with ${\Omega=\bigoplus_{k=1}^{n} i \sigma_{y}}$ with $\sigma_{y}$ being the Pauli $y$ matrix, $\partial_j$ is the derivative over the $j$-th unknown parameter, and repeated indices imply summation. The QFI of estimating $\theta,|g|$ is derived as (we use $\theta,|g|$ to label the corresponding matrix element of the QFI)
\begin{widetext}
\begin{equation}
\begin{aligned}\label{FI}
&F_{\theta\theta}=\frac{2\epsilon^2|g|^2}{2y+\epsilon(2+\epsilon-\epsilon|g|^2+2y)}\\
&F_{|g||g|}=\frac{2\epsilon^2[-\epsilon(2+\epsilon)^2+\epsilon^3|g|^4-4(1+\epsilon)(2+\epsilon)y-4(2+\epsilon)y^2]}{[\epsilon(-1+|g|^2)-2y][\epsilon(-2-\epsilon+\epsilon|g|^2)-2(1+\epsilon)y][\epsilon^2(-1+|g|^2)-4(1+y)-2\epsilon(2+y)]}\\
&F_{\theta|g|}=0,
\end{aligned}    
\end{equation}
\end{widetext}
where the squeezing level $y=2e^{-2r'}$ is used to quantify the amount noise introduced in the teleportation protocol with finite squeezing, which vanishes in the infinite squeezing limit, i.e., $\lim_{r\rightarrow \infty }y=0$.\\
\indent The QFI as a function of squeezing level $y=2e^{-2r'}$ is plotted in Fig.~\ref{qfi}. 
Although our scheme works for sources of arbitrary strength, it is important to check its performance in the weak limit $\epsilon\ll 1$. This is because in the weak limit, the estimation of the coherence function is strongly affected by vacuum noise, which degrades sensitivity if there is no shared entanglement between the two telescopes and only local measurements are performed at each telescope, as pointed out in Ref.~\cite{tsang2011quantum}. Reference \cite{tsang2011quantum} shows a local scheme without entanglement will at most have a Fisher information of $F\propto \epsilon^2$. To provide a more quantitative comparison, we now look at the resultant QFI of our scheme with squeezing parameter $r$ in different limits:\\
\textit{Low squeezing limit} with $y=2e^{-2r'}\rightarrow 2$:
\begin{equation}
\begin{aligned}\label{FI}
&F_{\theta\theta}\stackrel{\text{$y\rightarrow 2$}}{\Rightarrow}\frac{2\epsilon^2|g|^2}{4+\epsilon(6+\epsilon-\epsilon|g|^2)}\stackrel{\text{$\epsilon \ll 1$}}{\Rightarrow}\frac{1}{2}|g|^2\epsilon^2\\
&F_{|g||g|}\stackrel{\text{$y\rightarrow 2,\epsilon \ll 1$}}{{\Longrightarrow}}\frac{1}{2}\epsilon^2,
\end{aligned}    
\end{equation}
which coincides with the performance of heterodyne detection of a weak thermal source in telescopes \cite{tsang2011quantum}. Actually, the condition of low squeezing implies that the two-mode squeezed state becomes the vacuum state used in the protocol, since a local heterodyne measurement is performed on Telescope A and the corresponding QFI of the conditional state on Telescope B will be saturated by applying a heterodyne measurement therein as well. A similar observation has been made independently by \cite{Huang2024}; however, they used a slightly different CV-teleportation protocol, as we will explain later. \\
\textit{High squeezing limit} with $y=2e^{-2r'}\ll \epsilon$:
\begin{equation}
\begin{aligned}\label{FI}
&F_{\theta\theta}\stackrel{\text{$y\ll \epsilon $}}{{\Rightarrow}}\frac{2\epsilon|g|^2}{2+\epsilon-\epsilon|g|^2}\stackrel{\text{$\epsilon \ll 1 $}}{{\Rightarrow}}\epsilon |g|^2\\
&F_{|g||g|}\stackrel{\text{$y\ll \epsilon $}}{{\Rightarrow}}\frac{2\epsilon[2+\epsilon(1+|g|^2)]}{[1-|g|^2][\epsilon^2(1-|g|^2)+4+4\epsilon]}\\
&\quad\quad~~\stackrel{\text{$\epsilon \ll 1 $}}{{\Rightarrow}}\frac{\epsilon}{1-|g|^2},
\end{aligned}    
\end{equation}
which shows our scheme has $F\propto \epsilon$. Since $\epsilon\ll 1$ and the inverse of $F$ is the lower bound for the variance of the estimation, we will demonstrate below that our scheme can significantly outperform local schemes that do not utilize entanglement.

\yk{We now consider the optimal measurement that attains the sensitivity limit set by the QFI. To ensure optimality of the  identified  measurement, we allow the measurement to be any positive operator-valued measure (POVM), a broad framework that extends beyond projective measurements and provides greater flexibility for describing realistic measurement strategies \cite{nielsen2002quantum}. Quantum estimation theory provides a systematic approach to identifying the optimal POVM that achieves the sensitivity limit, utilizing the symmetric logarithmic derivative (SLD) \cite{braunstein1994statistical,paris2009quantum}. However, as we will see, the optimal measurement prescribed by quantum estimation theory is actually a simple projective measurement. } Since the teleported state is a Gaussian state, we can determine the SLD as a function of mean value $x_i$ and covariance matrix $\Sigma$ \cite{monras2013phase,gao2014bounds} as follows:
\begin{equation}\label{SLD_formula}
\mathcal{L}_{i}=\frac{1}{2} \mathfrak{M}_{\gamma \kappa, \alpha \beta}^{-1}\left(\partial_{i} \Sigma_{\alpha \beta}\right)\left(\mathbf{a}_{\gamma} \mathbf{a}_{\kappa}-\Sigma_{\gamma \kappa}\right),
\end{equation}
where we sum over repeated indices. For the estimation of $|g|$, the SLD is
\begin{equation}\label{SLDg}
\begin{aligned}
\mathcal{L}_{|g|}&=\frac{a}{d}(2\hat{a}_4^\dagger \hat{a}_4+1)+\frac{c}{d}(2\hat{a}_2^\dagger \hat{a}_2+1)\\
&\quad\quad+\frac{2b}{d}\hat{a}_4\hat{a}_2^\dagger+\frac{2b^*}{d}\hat{a}_4^\dagger \hat{a}_2+e,
\end{aligned}
\end{equation}
where the coefficients $a,b,c,d,e$ are given in Appendix~\ref{appendix:coefficients}. 
To find its eigenbasis, we define $\hat{d}_1=(\hat{a}_4+e^{i\delta}\hat{a}_2)/\sqrt{2}$, $\hat{d}_2=(\hat{a}_4-e^{i\delta}\hat{a}_2)/\sqrt{2}$:
\begin{equation}
\begin{aligned}
\mathcal{L}_{|g|}&=\hat{d}_1^\dagger \hat{d}_1(a+c+be^{i\delta}+b^* e^{-i\delta})/d\\
&\quad\quad+\hat{d}_2^\dagger \hat{d}_2(a+c-be^{i\delta}-b^* e^{-i\delta})/d\\
&\quad\quad+e+\frac{a+c}{d}\\
&\quad\quad+\hat{d}_1^\dagger \hat{d}_2 (a-c+be^{i\delta}-b^* e^{-i\delta})/d\\
&\quad\quad+\hat{d}_2^\dagger \hat{d}_1 (a-c-be^{i\delta}+b^* e^{-i\delta})/d.
\end{aligned}
\end{equation}
We want to choose $\delta$ such that $(a-c+be^{i\delta}-b^* e^{-i\delta})/d=(a-c-be^{i\delta}+b^* e^{-i\delta})/d=0$, which implies the eigenbasis is the Fock basis of $d_{1,2}$ modes. If $y\ll 1$ and $y\ll \epsilon$, to the leading order of $y$, we have 
\begin{equation}
\begin{aligned}
&(a-c+be^{i\delta}-b^* e^{-i\delta})/d\\
&=-(a-c-be^{i\delta}+b^* e^{-i\delta})/d\\
&=\frac{2i\sin(\delta-\theta)(2+\epsilon+\epsilon|g|^2)}{(-1+|g|^2)[-4-4\epsilon+\epsilon^2(-1+|g|^2)]}+o(1).
\end{aligned}
\end{equation}
This means we can choose $\delta=\theta$ in the measurement, and the POVM is $\{\ket{m,n}\bra{m,n}\}_{m,n}$, where $d_1^\dagger d_1\ket{m,n}=m\ket{m,n}$ and $d_2^\dagger d_2\ket{m,n}=n\ket{m,n}$. This measurement can be implemented using beam-splitters and photon-number-resolving detection with phase delay $\delta=\theta$, as shown in Fig.~\ref{set_up}.

For the estimation of phase $\theta$, the SLD is given by
\begin{equation}
\begin{aligned}
\mathcal{L}_{\theta}&=2p^* a_4a_2^\dagger+2pa_4^\dagger a_2,
\end{aligned}
\end{equation}
where the coefficient $p$ is given in Appendix~\ref{appendix:coefficients}.
To find its eigenbasis, we define $d_1=(a_4+e^{i\delta}a_2)/\sqrt{2}$, $d_2=(a_4-e^{i\delta}a_2)/\sqrt{2}$,
\begin{equation}\label{SLDtheta}
\begin{aligned}
\mathcal{L}_{\theta}&=d_1^\dagger d_1(p^*e^{i\delta}+p e^{-i\delta})-d_2^\dagger d_2(p^*e^{i\delta}+p e^{-i\delta})\\
&+d_1^\dagger d_2 (p^*e^{i\delta}-p e^{-i\delta})+d_2^\dagger d_1 (-p^*e^{i\delta}+p e^{-i\delta}).
\end{aligned}
\end{equation}
To have $(p^*e^{i\delta}-p e^{-i\delta})=0$, we can choose $\delta=\theta+\pi/2$. And the POVM is $\{\ket{m,n}\bra{m,n}\}_{m,n}$, where $d_1^\dagger d_1\ket{m,n}=m\ket{m,n}$ and $d_2^\dagger d_2\ket{m,n}=n\ket{m,n}$ (with a different $\delta$ when compared with the estimation of $|g|$). This POVM can be also implemented using beam-splitters and photon-number-resolving detection as in Fig.~\ref{set_up} with phase delay $\delta=\theta+\pi/2$. 

Since in the limit $y\ll \epsilon$, we can expect the teleported light to be the same as the state after directly bringing the light together to one location losslessly, the classical measurement schemes developed in astronomy should work as usual. For example, if we just measure the intensity difference between the two output ports $d_{1,2}$ with observable $\hat{O}=(d_1^\dagger d_1-d_2^\dagger d_2)/(2\epsilon)$, we have $\langle\hat{O}\rangle=|g|\cos(\theta-\delta)$ and the variance 
\begin{equation}\begin{aligned}
&\langle\hat{O}^2\rangle-\langle\hat{O}\rangle^2\\
&=[\epsilon+y+\epsilon(\epsilon/2+y)-\epsilon^2|g|^2/2+\epsilon^2|g|^2\cos^2(\theta-\delta)]/\epsilon^2,
\end{aligned}\end{equation}
which is comparable to the performance predicted by the QFI. Note that we do not need to know the squeezing parameter $r$ to estimate the coherence function and this estimator is unbiased even if the squeezing level is finite or even lower than the threshold. Intuitively, when $y$ is finite, there is background noise introduced by $y$ for the intensity measured at $d_{1,2}$, which is canceled when we take the difference between the intensities at $d_{1,2}$.

Furthermore, since $\theta$ is an unknown parameter, it is in general not possible to implement the optimal measurement strategy for the estimation of $\theta$ and $|g|$. We can certainly do the measurement adaptively and gradually optimize the phase in practice. We here want to briefly discuss the influence of having a phase delay in our measurement deviating from the optimal values. For simplicity, we assume $y\ll\epsilon\ll 1$. And we again consider the measurement that interferes with the light after adding the phase delay $\delta$ just as in Fig.~\ref{set_up}. The Fisher information (FI) of estimating $\theta$ and $g$ using this measurement to the leading order of $\epsilon$ is
\begin{equation}
\begin{aligned}
&I_{\theta\theta}=\epsilon\frac{|g|^2\sin^2(\theta-\delta)}{1-|g|^2\cos^2(\theta-\delta)},\\
&I_{|g||g|}=\epsilon\frac{\cos^2(\theta-\delta)}{1-|g|^2\cos^2(\theta-\delta)},\\
&I_{\theta |g|}=-\epsilon\frac{|g|\sin(\theta-\delta)\cos(\theta-\delta)}{1-|g|^2\cos^2(\theta-\delta)}.
\label{eq:FI}
\end{aligned}
\end{equation}
We observe that the optimal $\delta$ for estimating $\theta$ is $\delta=\theta+\pi/2$ and the optimal $\delta$ for estimating $|g|$ is $\delta=\theta$. The performance  decreases if $\delta$ deviates from the optimal values $\delta=\theta$, $\theta+\pi/2$ for estimation of $\theta$ and $|g|$, respectively. If we focus on scenarios where the size of the source exceeds the resolution limit and the intensity distribution becomes more complex than simple point sources, it is highly probable that $|g| \ll 1$. This is because the parameter $g$ represents a weighted sum of various phases. When these weights approach a more uniform or flat distribution, $g$ tends to approach 0. So, the performance is mainly determined by the numerator of the FI in Eq.~\ref{eq:FI}. So, reasonable performance can still be achieved as long as the deviation of $\delta$ from the optimal choice is not too large. 

\section{Comparison with other proposals for astronomical interferometers}
\label{sec:compare}

\yk{We first want to expand on the practical implications of our method  based on CV teleportation  for astronomical imaging. The resolution of an imaging system is fundamentally determined by the diffraction limit, given by $\lambda/d$, where $\lambda$ is the observation wavelength and $d$ is the aperture diameter for single-lens imaging or the baseline length for interferometric imaging. The highest angular resolution in ground-based astronomical imaging has been achieved using radio interferometry, where the maximum baseline is limited by the diameter of the Earth. This limit has already been reached in current radio astronomical interferometers, such as the Event Horizon Telescope (EHT), which observed M87 at a wavelength of 1.3 mm \cite{EHT2019}. Given an Earth-sized baseline of approximately 13,000 km, the corresponding resolution is around 0.02 mas, where we express the resolution in milliarcseconds (mas) rather than radians.

A significant improvement in resolution can be achieved by reducing the observation wavelength. If similar interferometric techniques were applied at optical wavelengths—for example, at 1550 nm—the achievable resolution would improve to approximately $2.5 \times 10^{-5}$ mas, nearly two orders of magnitude finer than current radio interferometry.
However, existing optical astronomical interferometers, such as the CHARA array \cite{monnier2003optical}, have a baseline limited to around 330~m, significantly shorter than the Earth-sized baselines used in radio interferometry. Extending the baseline for optical astronomical interferometry is therefore a crucial challenge. A promising solution is the use of quantum-enhanced astronomical interferometry based on quantum networks, as initially proposed in Refs. \cite{gottesman2012longer, khabiboulline2019quantum, khabiboulline2019optical} and further developed in our work.

To put this into perspective, one of the smallest astronomical objects observed to date using a radio interferometer array is the central compact source of the supermassive black hole candidate at the core of the giant elliptical galaxy M87, with an angular size of 0.04 mas \cite{EHT2019}. In contrast, one of the smallest sources observed at optical wavelengths is exoplanet host stars, which typically have angular diameters of about 1 mas \cite{boyajian2015stellar}. This highlights the potential of high-resolution optical interferometry to resolve even finer details than what is currently achievable.

Moreover, observations at different wavelengths provide complementary information about astronomical sources. If optical interferometry could achieve resolutions comparable to or even surpass those of radio frequencies, it would become a powerful tool for high-resolution astronomical imaging, significantly enhancing our ability to study the fine details of celestial objects.

}

In the following, we compare our proposal with several other proposals for astronomical interferometers and elaborate the advantages as indicated by the FI. \yk{We first compare our proposal with the direct detection method, which involves transmitting the stellar light collected by two distant telescopes to a single location, where the stellar light interferes and is then measured.} The QFI/FI $F_{\theta\theta},F_{|g||g|}$ lower-bound the variance of estimating the coherence function $|g|,\theta$, which implies that the signal-to-noise ratio (SNR) for $|g|,\theta$ is lower-bounded by $\theta \sqrt{F_{\theta\theta}N}$, $|g|\sqrt{F_{|g||g|}N}$, where $N$ is the number of measurement events. So, the advantage identified for QFI/FI can be directly translated to the advantage for SNR in astrophysics. \yk{Consider the previously mentioned CHARA interferometer array \cite{monnier2003optical}, a currently operating optical astronomical interferometer with a baseline of approximately 330~m.  We provide a rough example by considering a loss of 1 dB/km as an imperfection in the practical implementation, which corresponds to the transmission loss of free space propagation in weakly degraded conditions 
\cite{pan2023free,bloom2003understanding,khalighi2014survey}. The total transmission loss is $T=0.92683$.} we can then calculate the SNR of estimating the phase of the coherence function $\theta$ as 
\begin{equation}
\text{SNR}_\theta=\frac{\theta}{\Delta\theta}\leq\theta \sqrt{N I_{\theta\theta}}=\theta|g| T\sqrt{N\epsilon},
\end{equation}
where $N$ is the number of temporal modes we measure and $I_{\theta\theta}$ is the FI of estimating $\theta$ using direct detection. We can set $|g|=10^{-2}$, $\theta=\pi/4$, $\epsilon=10^{-4}$, $N=2\times 10^8$; then $\text{SNR}_\theta\approx 1$. Note we roughly need $N\epsilon=2\times 10^4$ photons to get $\text{SNR}_\theta\approx 1$ using direct detection.
The resolution of the optical interferometer is determined by the length of baseline as $\lambda/d$, where $\lambda$ is the wavelength and $d$ is the baseline. The CHARA telescope has resolution 
\begin{equation}
\frac{2.2~\mu \text{m}}{330~\text{m}}\times 206265000~\text{mas}/\text{rad}=1.375~\text{mas}.
\end{equation}
If we increase the baseline 100 times, 
the resolution becomes 0.01375 mas but SNR becomes 0.00054 for the same set of parameters and the number of  measured temporal modes $N$. We can, of course, improve the SNR by having a longer observation time; i.e., increasing $N$, which, though, means higher cost. So, it is clear that, limited by the SNR, we cannot extend the baseline for conventional methods too much. However, with our method, assuming a quantum repeater operates effectively between two distant telescopes, i.e. provides the effective squeezing parameter $r'$ over the threshold, it can enhance the SNR or extend the baseline while keeping the SNR constant.  For our CV scheme with squeezing level above the required threshold, the required number of temporal modes is $N=1.72\times 10^8$ to achieve $\text{SNR}_\theta\approx 1$. In other words, with $N=1.72\times 10^8$ temporal modes, we can still get $\text{SNR}_\theta\approx 1$ as long as we can build an effective CV repeater between these distant telescopes.
In the following, we will provide an example to show the threshold for the necessary effective squeezing parameters. In the next section, we will discuss in detail quantum telescopes using CV-quantum teleportation in the absence and presence of quantum repeaters.

\begin{figure}[!tbh]
\begin{center}
\includegraphics[width=1\columnwidth]{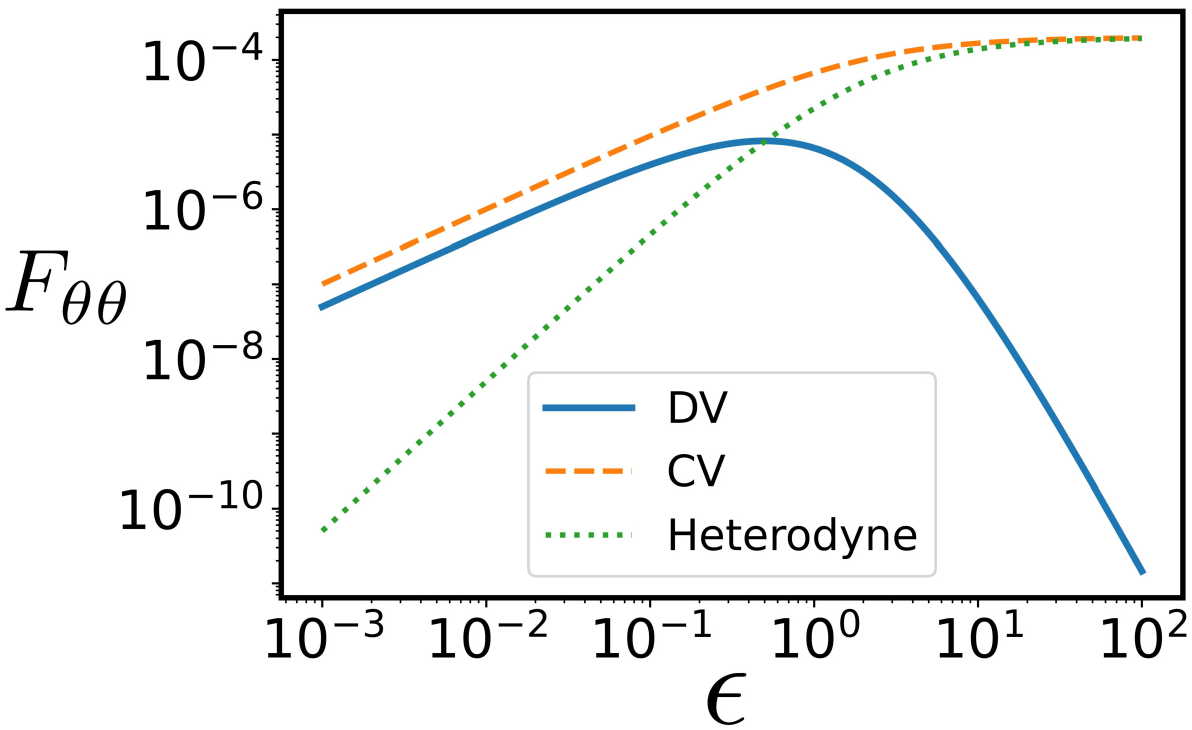}
\caption{The Fisher information (FI) for estimating the phase $\theta$ of the coherence function by discrete variable (DV) quantum-network-based astronomical interferometry, continuous variable (CV) quantum-network-based astronomical interferometry, and local heterodyne detection, as a function of the mean photon number per temporal mode of the stellar light $\epsilon$. 
}
\label{compare}
\end{center}
\end{figure}

We can also explicitly compare our scheme based on a CV quantum network in the high squeezing limit with the scheme based on discrete-variable (DV) quantum networks proposed by Ref.~\cite{gottesman2012longer} and heterodyne techniques by calculating the FI of estimating $\theta$. As shown in Fig.~\ref{compare}, as the mean photon number $\epsilon\ll 1$, we observe the ratio between the FI of the CV and DV methods approach a constant of $1/2$. This factor of 1/2 is due to the fact that Ref.~\cite{gottesman2012longer} wastes half of the stellar photons because it is not possible to implement the whole set of Bell measurements only using linear optics. In contrast, our scheme relies on homodyne detection instead and can be implemented with only linear optics. As $\epsilon$ increases, the performance gap between the schemes based on DV and CV quantum networks increases. This is because multiphoton events become more and more important as $\epsilon$ increases, and since our method makes use of the general stellar state with possibly more than one photon in Eq.~\ref{rho}, it can use multiphoton events instead of regarding them as noise as in Ref.~\cite{gottesman2012longer,khabiboulline2019optical,khabiboulline2019quantum}. 
As an example, consider a stellar source with $\lambda=800~\text{nm}$, a bandwidth $\Delta\lambda=0.1~\text{nm}$, and a magnitude of -1.43 (the magnitude of the brightest star, Sirius), observed by telescopes with an 18 m diameter, which have $\epsilon \approx 0.4$. 
For $|g|=0.01$ and $\theta=\pi/4$, the FI of the CV method is $F_{\theta\theta}=3.33\times 10^{-5}$, while for the DV method, it is $F_{\theta\theta}=8.04\times 10^{-6}$. This indicates that the variance in estimating $\theta$ for our CV method will outperform the DV method by a factor of 4.147 with the same number of temporal modes. The factor 4.147 arises from two contributions as pointed out above. First, the DV teleportation performs worse by a factor of 1/2 because a full Bell measurement cannot be implemented only using linear optics. Second, multi-photon events from a relatively strong source can be measured in the CV telescope but are ignored in the DV telescope.
To achieve the same $\text{SNR}_\theta \approx 1$, our CV method would require approximately $5\times 10^4$ temporal modes, whereas the DV method would need around $2\times 10^5$ temporal modes.
The above calculation of the FI for the DV method assumes a direct implementation of Ref.~\cite{gottesman2012longer}. It is possible to bridge this performance gap in the DV method using more entanglement by performing teleportation in higher dimensions, which is generally more challenging to implement, as discussed in Ref.~\cite{luo2019quantum}. In contrast, the continuous-variable (CV) scheme has a relatively lower threshold for the squeezing parameter when imaging brighter sources as estimated using $2e^{-2r}\sim \epsilon$, which can be as low as 7~dB in our example.
The estimation of $\epsilon$ may vary depending on the parameters chosen. However, in general, the brightest source imaged by large telescopes might be within the regime where our scheme shows advantages over the DV scheme.


 However, we emphasize that as $\epsilon$ becomes large enough, it becomes possible to perform measurements locally without any entanglement resources  as claimed in Ref.~\cite{tsang2011quantum}. We calculate the performance of estimating $\theta$ only using heterodyne detection locally at each telescope (the FI calculation for heterodyne detection is given in Appendix~\ref{Appendix:heterodyne}). It is clear from the figure that for $\epsilon\gg 1$, local heterodyne detection without any entanglement can perform as well as our scheme based on a CV quantum network for the estimation of phase $\theta$. This shows that there exist schemes that only do measurement locally without any entanglement and still have the optimal performance for imaging strong thermal sources using interferometry with two telescopes. For intermediate values of $\epsilon$, our method can perform significantly better than methods based on DV quantum networks and local heterodyne detection.

When comparing to local heterodyne detection, our protocol
offers significant advantage in the weak source limit, as shown in Fig.~\ref{compare}. This contrasts with the findings reported in Ref.~\cite{Huang2024}, where it is argued that the quantum enhancement using two-mode squeezed vacuum states and CV-teleportation is limited.
We would like to emphasize that this conclusion about minimal enhancement is drawn from scenarios lacking a CV repeater connecting distant telescopes. In cases without a CV repeater, the two-mode squeezed state source used for teleportation shows only a low effective squeezing level, which is consistent with the low squeezing limit of our scheme. However, with a well-performing CV repeater, the potential benefits could be substantial. Furthermore, a close comparison with our protocol reveals that the key difference between these two CV-teleportation based protocols is the presence and absence of a conditional displacement operation. In our protocol, we introduce the conditional displacement operation as shown in Fig.~\ref{set_up} such that the measurement at telescope B is performed after completion of the whole CV-teleporation protocol, while a post-selected teleportation is performed in Ref.~\cite{Huang2024} and the quantum Fisher information of the teleported state at telescope B is post-processed and computed from the ensemble average over different measurement outcomes at telescope A. 
An additional difference between the schemes is that, in order to perform the conditional displacement operation in our scheme, classical information is required to be sent from telescope A to telescope B, so, in general, a delay line needs to be included locally at telescope B for synchronization. Such a delay line could be made much more resilient to noise and loss in the lab compared to the optical channel connecting two telescopes, for example, using a broadband all-optical delay-line quantum memory \cite{Nathan2024}.

\section{Amplification of entanglement resources}

\begin{figure}[!tbh]
\begin{center}
\includegraphics[width=1\columnwidth]{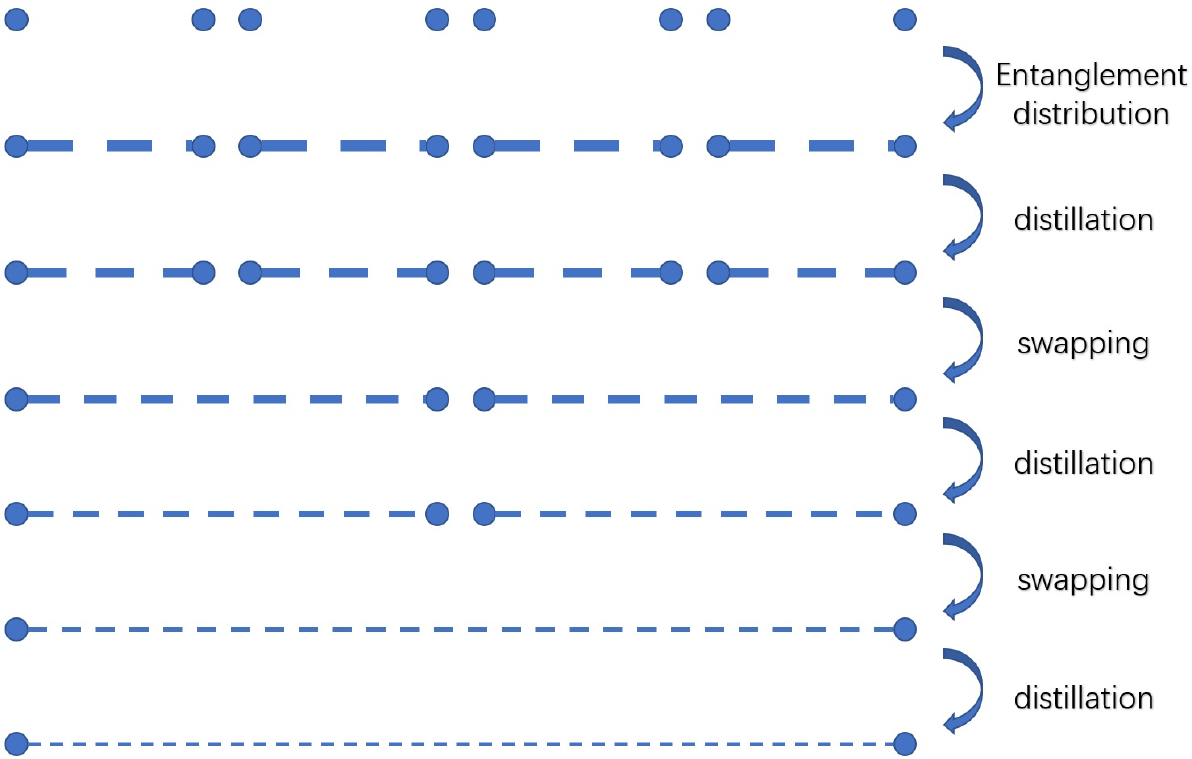}
\caption{Continuous-variable quantum repeater. The two-mode squeezed states are distributed to adjacent quantum nodes through lossy channels. After several rounds of entanglement distillation and entanglement swapping, we create entanglement over a long distance. The blue circles are quantum nodes. The dashed line indicates entanglement and the thickness indicates the number of copies. The states are stored in quantum memories in each quantum node until their use.}
\label{repeater}
\end{center}
\end{figure}

\begin{figure*}[!tbh]
\begin{center}
\includegraphics[width=1.7\columnwidth]{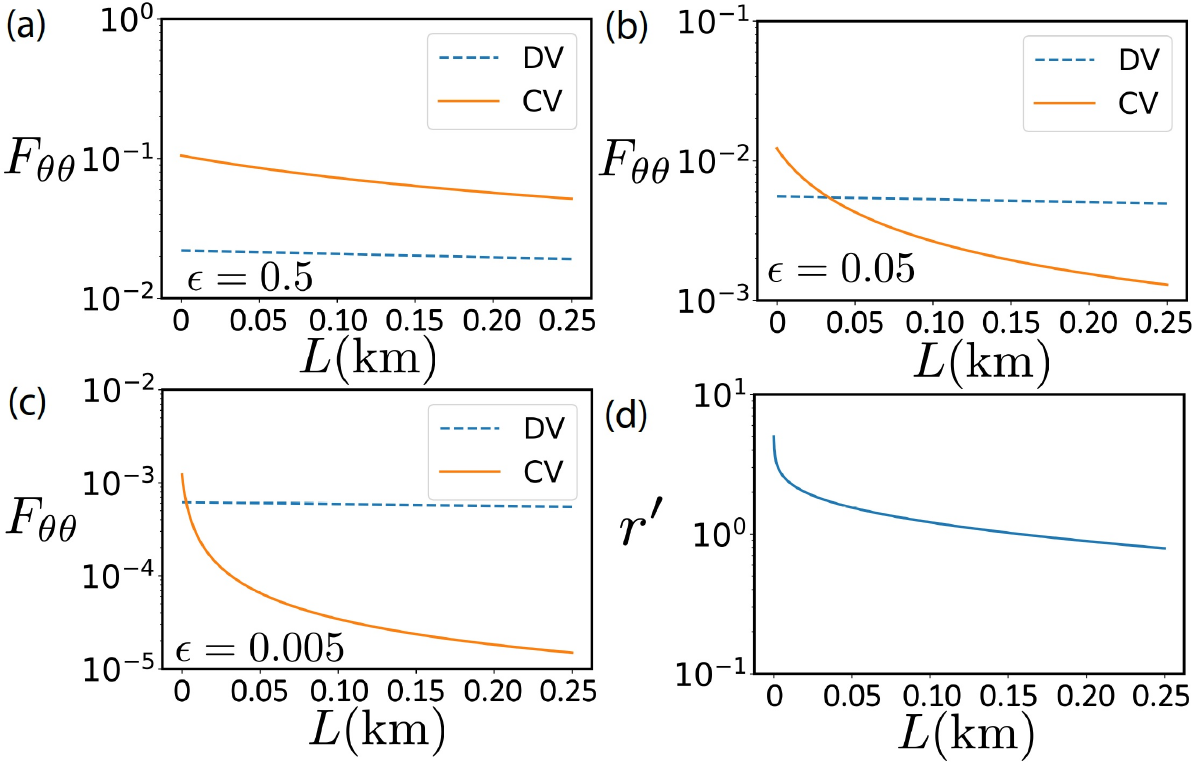}
\caption{Influence of transmission loss on interferometric imaging based on discrete-variable (DV) and continuous-variable (CV) schemes when there is no entanglement distillation or quantum repeaters. We assume the transmission loss is 0.2 dB/km. (a-c) The Fisher information $F_{\theta\theta}$ of estimating the phase $\theta$ of the coherence function versus the baseline between telescopes $L$, for the mean photon number per temporal mode $\epsilon=0.5,0.05,0.005$. For the CV case, we assume the squeezing parameter $r=5$ before the two-mode squeezed state is sent through the lossy channel. (d) The effective squeezing parameter $r'$ decreases with the baseline $L$.}
\label{loss}
\end{center}
\end{figure*}

We first explore for both DV and CV cases the situation where entanglement resources are distributed without any entanglement distillation or quantum repeaters. As shown in Fig.~\ref{loss}, we consider the Fisher information (FI) of estimating the phase of the coherence function versus the baseline length $L$ for different strengths $\epsilon$ of the stellar source. \yk{
We assume a typical transmission loss of 0.2 dB/km for optical fiber at 1550 nm  \cite{keiser2000optical, hasegawa2018first}.}
The FI for the DV case decays exponentially with $L$ because loss increases for the terrestrial photons, resulting in a higher probability for vacuum state contributions. The behavior of the FI as a function of $L$ in the CV cases depends on the source strength $\epsilon$. This is because the value of the effective squeezing parameter $r'$ at which the QFI starts to significantly drop depends on $\epsilon$. This can be seen from Fig.~\ref{qfi} and Eq.~\ref{FI}, which show $y=2e^{-2r'}$ should be at least comparable to $\epsilon$ to obtain sensitivity close to the optimal case when $y\rightarrow 0$. To make $r'$ comparable to $r$, the effective transmission coefficient $T$ needs to satisfy $1-T^2\sim e^{-2r}$. Given a squeezing parameter $r$ high enough such that the QFI approaches its optimal value, to keep reasonable performance $1-T^2$ should be at least comparable to $\epsilon$. When $\epsilon$ is small, the requirement on $T$ is very high and the performance of our scheme based on CV teleportation decays very fast. When $\epsilon\sim O(1)$, our scheme based on CV teleportation becomes more robust to loss. This is because in this case, the threshold of $T$ can be estimated by $1-T^2\sim \epsilon\sim O(1)$, which suggests a relatively relaxed requirement of $T \sim O(1)$.

An important question  is how to achieve in practice a high-fidelity two-mode squeezed state shared by distant telescopes connected by lossy quantum channels. As discussed in the previous section, a lossy squeezed state will strongly limit the sensitivity. As in the discrete case \cite{gottesman2012longer} where one must build shared entanglement between distant telescopes with a DV quantum repeater, we will need to 
apply the  CV quantum repeater to the problem of creating long-baseline astronomical interferometers \cite{ralph2009nondeterministic,ralph2011quantum,dias2017quantum,furrer2018repeaters,liu2018continuous,dias2020quantum,seshadreesan2020continuous}. This will allow us to extend the baseline of interferometers beyond what is possible to achieve by direct interference. 


\yk{A CV quantum repeater network, as illustrated in Fig.~\ref{repeater}, consists of a series of repeater nodes strategically placed at moderate distances from one another. The process of establishing entanglement between distant telescopes begins with the distribution of two-mode squeezed states between adjacent nodes. These initially noisy entangled states undergo entanglement distillation, which reduces noise by transforming a larger number of low-quality entangled states into a smaller set of higher-fidelity ones. Following this, entanglement swapping is employed to extend entanglement across multiple nodes, ultimately creating long-distance entangled links between telescopes \cite{dias2017quantum,furrer2018repeaters,dias2020quantum,seshadreesan2020continuous}.  In our setup for a quantum-enhanced astronomical interferometer, we avoid using long-lived quantum memories at the telescopes. This is in contrast to the memory-based quantum telescopy protocols proposed in \cite{khabiboulline2019optical, khabiboulline2019quantum}, which require a quantum memory with a lifetime exceeding $1$~ms for an experiment with $10$~GHz detector bandwidth and $\epsilon=10^{-7}$ source photon rate. A quantum repeater protocol typically demands  a quantum memory with only $\sim10~\mu$s lifetime for telescopes separated by a few kilometers. Moreover, it can even be implemented using an all-optical quantum repeater protocol, eliminating the need for quantum memory altogether \cite{Azuma2023}. 

CV quantum repeaters help mitigate two primary types of noise. The first type is the phase noise caused by the variation of the path length of the interferometer. This can be solved by active stabilization \cite{foreman2007remote}, where the distance is tracked by a reference laser. Or the phase noise can be overcome by entanglement distillation \cite{liu2018continuous}. The second major source of noise is transmission loss, which can be effectively reduced using entanglement distillation techniques.

One notable entanglement distillation method for overcoming transmission loss is based on nondeterministic noiseless linear amplification (NLA) \cite{ralph2009nondeterministic,ralph2011quantum}. Consider the distribution of two-mode squeezed states with a squeezing parameter $r$ through a channel with a specific transmission coefficient. In this method, the transmitted quantum state is first split into $N$ modes using balanced beam splitters. Each mode undergoes the quantum scissor operation \cite{ozdemir2001quantum,pegg1998optical}, which relies on linear optics and single-photon detection. Conditioned on successful detection events, the quantum scissor truncates the state to the Fock subspace ${\ket{0}, \ket{1}}$ while simultaneously amplifying the coefficient of $\ket{1}$. The modes are then recombined using a beam splitter and post-selection is performed by detecting vacuum in all but one output port. This process yields an amplified two-mode squeezed state with an improved effective squeezing parameter  and higher effective transmissivity, though it comes at the cost of probabilistic success due to the reliance on specific measurement outcomes. This technique enables the distillation of entanglement between distant nodes, allowing highly entangled states to be established even in the presence of significant transmission loss.

Other entanglement distillation methods for combating transmission loss include symmetric photon replacement \cite{lund2009continuous} and purifying distillation \cite{fiuravsek2010distillation}. These non-Gaussian operations generally produce non-Gaussian states as outputs, which can be further processed using Gaussification protocols to restore Gaussian characteristics when needed \cite{browne2003driving,eisert2004distillation}.

While the development of CV quantum repeaters is still in its early stages, ongoing research is exploring ways to enhance their performance. One promising direction involves the integration of CV quantum error correction codes \cite{rozpkedek2021quantum,lassen2010quantum,lassen2013gaussian}, which could form the foundation of  quantum repeaters, offering improved robustness against noise and loss. 
Exploring the potential role of polarization-squeezed states in CV quantum teleportation and repeater protocols represents an intriguing future direction \cite{lassen2007generation}, while quantum averaging offers techniques for stabilizing squeezed states and could be valuable for the future development of CV quantum repeaters \cite{lassen2010experimental}.
For more details regarding the implementation of CV quantum repeaters,  please refer to Appendix~\ref{appendix_repeater}.

}

For the purpose of imaging, the parameters relevant to performance are squeezing parameter $r$, transmission loss in the distribution of entanglement $T$, and repetition rate. A CV quantum repeater can ensure transmission loss is limited to a small constant over long distances \cite{dias2017quantum}, which suggests a fixed $T$ in the output of the repeater. 
Recall that we have defined effective squeezing parameter $r'$ such that $e^{-2r'}=1-T^2+e^{-2r}T^2$. As a simple model, we consider the case of fixing $r'$  with the help of CV repeaters even when the distance increases. Then, the goal is to make sure $T$ is large enough so that $r'$ can be comparable to $r$. The repetition rate of a quantum repeater depends on the rate of the sources at each repeater node and the success probability of quantum operations at each node, such as entanglement distillation and swapping. 
The repetition rate will be polynomial in the distance between two telescopes. If we consider the CV repeater proposed in Ref.~\cite{dias2017quantum}, the transmission loss between nearby repeater nodes and the squeezing level will determine the order of the polynomial.

 As the squeezing level and repetition rate of the quantum repeater become high enough, we can expect the quantum teleportation to be approximately ideal. In the previous section, we examined the threshold for the squeezing parameter using a specific example involving very bright sources and large telescopes. In the following section, we will provide a more comprehensive analysis of this threshold, considering smaller telescopes and varying magnitudes of sources. If we choose $\lambda=800\;\text{nm}$, $\Delta\lambda=0.1\;\text{nm}$, the threshold for the repetition rate of our repeater to make sure our distributed two-mode squeezed states cover all the temporal modes is roughly 150~GHz. We assume the diameter of each telescope is 6~m. We list the threshold squeezing level for different magnitudes of stellar sources in Table~\ref{squeezing_table}. 
 If we can distribute two-mode squeezed states with both the squeezing level and repetition rate better than their thresholds, our scheme can overcome the transmission loss and hence significantly outperform the direct detection case which suffers from the transmission loss. Since $\epsilon$ will determine the required squeezing level and very strong squeezing levels are required to image weak sources, it may be advantageous to increase the mean photon number per temporal mode by having more than two telescopes work together in this scheme while imaging weak sources. But this scheme would require a modified version of the standard CV quantum teleportation, which is left as a possible future direction of work. 
 

\yk{We now seek to further assess the technical feasibility of implementing our scheme. Building an optical parametric oscillator (OPO) with nonlinear crystals is a well-established method for generating squeezed light. Recent experiments using continuous-wave OPOs have achieved squeezing levels as high as 15 dB below the vacuum noise limit \cite{vahlbruch2016detection}. However, as the level of enhancement increases, the operational bandwidth becomes increasingly narrow. An alternative approach leverages the $\chi^{(3)}$-nonlinearity of optical fibers, where intense laser pulses propagate through low-loss, highly nonlinear fibers to induce squeezing without requiring the light to resonate in a cavity. Although the squeezing levels obtained with this method are generally lower than those achieved with OPOs \cite{andersen201630}, higher repetition rates can be realized using a pulsed light source. For instance, a repetition rate of 163 MHz combined with 6.8 dB of squeezing has been demonstrated using fiber-based techniques \cite{dong2008experimental}. A third method employs optical waveguides to tightly confine light spatially. By combining waveguides with pulsed light sources, it is, in principle, possible to generate broadband, highly squeezed light with enhanced compactness and stability;  an experiment reported a repetition rate of 86.5 MHz and a squeezing level of 5.88 dB \cite{amari2023highly}.

Although the ambitious target of 150 GHz repetition rate remains a technical challenge, advances in material engineering and waveguide design are paving the way towards this goal. Furthermore, as shown in Table~\ref{squeezing_table} and Fig.~\ref{compare}, in the regime of $\epsilon \sim 1$—where our CV scheme outperforms both DV methods and local heterodyne detection—the threshold squeezing level is within the capabilities of current state-of-the-art technology. While this reflects only the current capabilities of squeezed state generation, realizing our approach will also require quantum repeaters to distribute the entanglement and distill noisy two-mode squeezed states. Nonetheless, these developments highlight the potential feasibility of implementing our method. 
}

\begin{figure}[!tbh]
\begin{center}
\includegraphics[width=1\columnwidth]{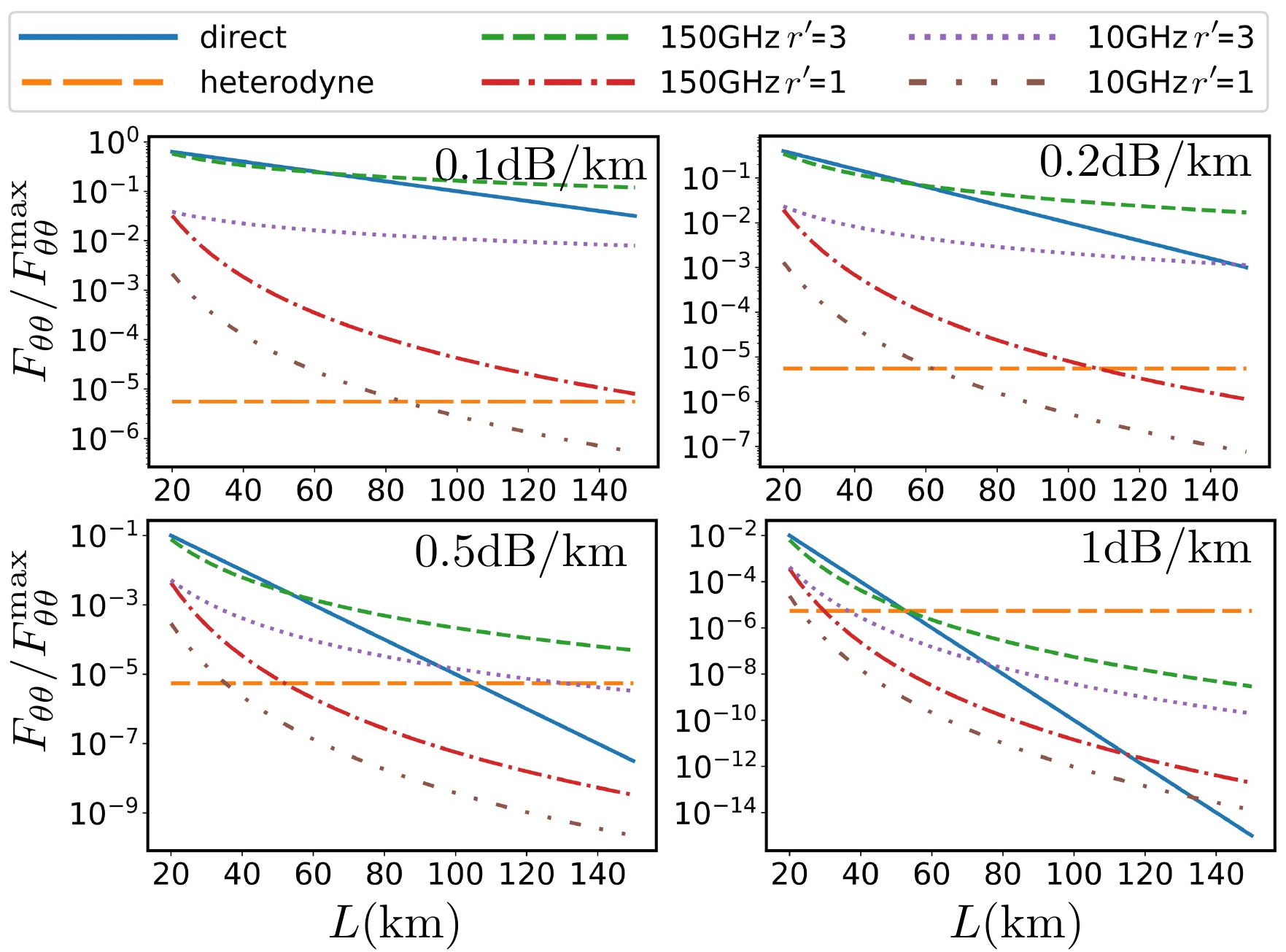}
\caption{\yk{Ratio of achieved FI over the maximum FI in the ideal case for estimating the phase of the coherence function $\theta$ for different transmission losses. The solid blue line corresponds to the case of directly combining the light from telescopes with baseline $L$ in units of kilometers. The horizontal dashed line corresponds to local heterodyne detection.   The other plotted lines are the performance of our CV quantum-network-based scheme for different repetition rates and effective squeezing parameters $r'$. Here we assume the quantum repeater operates with the chosen repetition rate at a distance of 10~km, and that the repetition rate starts to decrease polynomially with distances larger than 10~km. We assume $\lambda=800\;\text{nm}$, $\Delta\lambda=0.1\;\text{nm}$, the source has magnitude -5, and we are using telescopes of diameter 6~m.}}
\label{distance}
\end{center}
\end{figure}

\yk{We explicitly compare the performance of our scheme with both the direct-detection method—where light from two distant telescopes is directly combined—and local heterodyne detection at each telescope, as shown in Fig.~\ref{distance}. We focus on a relatively bright stellar source of magnitude -5, as our scheme demonstrates an advantage over both the DV scheme in Ref.~\cite{gottesman2012longer} and local heterodyne detection when $\epsilon \sim O(1)$, as previously discussed in  Fig.~\ref{compare}. For an astronomical source of magnitude -5 observed with 6-meter-diameter telescopes at $\lambda = 800~\text{nm}$ and spectral bandwidth $\Delta\lambda = 0.1~\text{nm}$, the mean photon number per temporal mode is approximately $\epsilon \sim 0.4$. This lies in the regime where the requirement for effective squeezing is moderate, with a threshold effective squeezing parameter of approximately $r' \approx 0.8$. The required repetition rate to cover all temporal modes is about 150~GHz.
We analyze a range of values for $r'$ and the repetition rate for quantum repeaters operating over a 10 km distance in Fig.~\ref{distance}. For longer distances, we account for the polynomial decrease in repetition rate while keeping $r'$ fixed, following the scaling in Ref.~\cite{dias2017quantum}. In practice, this can be achieved through entanglement distillation, which consumes multiple copies of noisy two-mode squeezed states to generate states with a larger effective squeezing parameter $r'$.

As shown in Fig.~\ref{distance}, even if the required repetition rate and squeezing values are not fully achieved, our scheme remains preferable as the distance increases. A key observation from this figure is that the performance of the direct detection method decreases exponentially with distance, whereas our scheme, based on CV quantum teleportation, exhibits only a polynomial decrease. Notably, this polynomial scaling persists even when the required repetition rate and squeezing parameter are not fully met, allowing for a parameter regime where our scheme outperforms direct detection even under non-ideal conditions. This highlights the feasibility of our approach despite technological challenges.

When comparing our scheme with the direct detection method under different transmission loss conditions, we find that direct detection can outperform our approach at short distances, particularly when the transmission loss is small. This is because the non-ideal implementation of our method introduces additional noise into the state. However, as the distance increases, the polynomial scaling of our scheme ensures that it eventually surpasses the direct detection method. Moreover, the transition distance $L$, beyond which our scheme becomes superior, decreases as the transmission loss increases. This is expected, as quantum repeaters are designed to mitigate transmission loss and should offer greater advantages in higher-loss scenarios.

The performance of local heterodyne detection depends only on the strength of the source and appears as a horizontal line in the figure. At sufficiently large distances, both the direct detection method and our CV quantum teleportation scheme eventually perform worse than local heterodyne detection. Furthermore, as transmission loss increases, the threshold distance at which heterodyne detection surpasses both methods decreases.

It is important to note that the performance of our scheme estimated here is based on a non-ideal model, where the repetition rate of the squeezing source is assumed to be limited. As the distance increases, we assume that the quantum repeater maintains the effective squeezing level by consuming noisy two-mode squeezed states, causing the repetition rate to decrease with distance. This demonstrative model, similar to that in Ref.~\cite{dias2017quantum}, captures the essential features of our scheme in practical settings. However, we emphasize that if a sufficiently high-performance CV quantum repeater can be implemented, our scheme will always outperform both the direct detection method and local heterodyne detection.

}

\begin{table}[h!]
  \begin{center}
    \caption{Threshold squeezing level for different magnitude of stellar sources with $\lambda=800\text{nm}$, $\Delta\lambda=0.1\text{nm}$. The threshold is calculated by setting $r'=-1/2\log(\epsilon/2)$. }
    \label{squeezing_table}
    \begin{tabular}{|c|c|c|c|} 
    \hline
      Magnitude & $\epsilon$ & Squeezing (dB) & $r'$\\
    \hline
    -5 & $4\times10^{-1}$ & 7 & 0.80\\
     \hline
    -2.5 & $4\times 10^{-2}$& 17 & 1.96\\
     \hline
    0 & $4\times 10^{-3}$ & 27 & 3.11\\
     \hline
     2.5 & $4\times 10^{-4}$ & 37 & 4.26\\
     \hline
     5 & $4\times 10^{-5}$ & 47 & 5.41\\
     \hline
     7.5 & $4\times 10^{-6}$ & 57 & 6.56\\
     \hline
    
    \end{tabular}
  \end{center}
\end{table}

We previously compared our scheme based on CV quantum repeaters with the scheme based on DV quantum repeaters proposed in Ref.~\cite{gottesman2012longer} in Fig.~\ref{compare}, assuming the repeaters are ideal. But the real-world implementation of CV and DV quantum repeaters will obviously affect the imaging quality of the two schemes. As the preliminary proposal for CV quantum repeaters appeared only recently \cite{dias2017quantum}, and the development of CV quantum repeaters is still in its early stages, it is hard for us to predict which scheme will eventually show better performance. We note there are some theoretical efforts on the direct comparison between DV and CV quantum repeaters \cite{dias2022distributing}. They consider the ability to distribute a two-mode squeezed vacuum as the figure of merit to compare DV and CV repeaters. Their results suggest that CV repeaters may outperform DV repeaters for a certain fidelity of DV entangled pairs. Although this work suggests an inspiring way to benchmark the two different approaches and suggests that CV repeaters may outperform DV repeaters in some parameter regimes, their comparison relies on several assumptions, such as using DV entanglement to distribute CV entanglement and the strength of the two-mode squeezed vacuum. The comparison between our scheme and Ref.~\cite{gottesman2012longer} in actual implementation requires more theoretical and experimental effort and is left as future work.


We now compare our method with the intensity interferometer \cite{brown1956correlation} and heterodyne interferometer \cite{hale2000berkeley}. The intensity interferometer measures the intensity fluctuations at two distant telescopes independently as a function of time, which is then  used to find the second-order correlation of the received light via data post-processing \cite{brown1956correlation}. For thermal light, the second-order correlation function is directly related to the coherence function we want to measure. Compared to our method, the intensity interferometer is much easier to implement since the measurement is independent in each telescope. But for weak thermal sources, the intensity interferometer has much worse sensitivity because it requires at least two photons within the same temporal mode to extract useful information, meaning many temporal modes with one photon are wasted. Heterodyne detection mixes the light from the astronomical source with a laser on a beam-splitter in order to measure the coherence function \cite{hale2000berkeley}. Although the length of the baseline will be limited due to the optical feedback system used for phase-locking lasers at the two telescopes, this is still much easier than distributing fragile entangled states. However, the sensitivity is once again worse than our method for weak sources. This is mainly because local heterodyne detection is unable to distinguish the vacuum state and states with at least one photon in the stellar light while measuring the coherence function, which introduces strong vacuum noise to the estimation, as pointed out in Ref.~\cite{tsang2011quantum}.

In gravitational wave detectors, which represent an important application of squeezed states, photon counting noise can be reduced by injecting the squeezed state into one port of the interferometer \cite{caves1981quantum}. A natural question is whether we can enhance the estimation of coherence using squeezed light instead of only using squeezed states as entanglement resources to perform quantum teleportation. The answer is unfortunately no. Intuitively, this is because in the gravitational wave detector \cite{caves1981quantum}, the squeezed state is also encoded with the unknown phase we want to estimate, i.e. the squeezed state can be regarded as part of the probe state. But for astronomical interferometers, the state we receive at the telescopes is already encoded with the information we want to measure. The squeezed state can only be used in ancillary modes and not as a probe state. More details can be found in Appendix~\ref{squeeze}.

\section{Conclusion}

In summary, we propose to use two-mode squeezed states as an entanglement resource to overcome transmission loss in astronomical interferometry. Our scheme is based on the CV teleportation of stellar light. The optimal measurement on the teleported states to estimate the coherence function is constructed using beam-splitters and photon-number-resolved detection. Due to phase noise and transmission loss in the distribution of the two-mode squeezed states, our scheme relies on CV quantum repeaters to build entanglement between distant telescopes.

\yk{After uploading the preprint of this work, we became aware of two subsequent independent studies conducted in a similar setting by Zixin Huang et al.~\cite{Huang2024} and Bran Purvis et al.~\cite{purvis2024practical}. The differences between their work and ours are discussed in Appendix~\ref{appendix:comparison}.}


\section{Acknowledgements}
We would like to  thank  Eric Chitambar, Andrew Jordan, Paul Kwiat, John D. Monnier, Shayan Mookherjea, Michael G. Raymer and Brian J. Smith  for helpful discussion. This work was supported by the multiuniversity National Science Foundation Grant No. 1936321 and No. 2326803.


\appendix
\section{Continuous variable quantum repeaters}\label{appendix_repeater}

In the basic CV quantum repeater, many copies of two-mode squeezed states are distributed to adjacent repeater nodes. As shown in Fig.~\ref{repeater}, each repeater node then does entanglement distillation and entanglement swapping \cite{dias2017quantum,furrer2018repeaters,dias2020quantum,seshadreesan2020continuous}. There are several types of entanglement distillation and swapping methods. We now introduce some representatives of them.

Entanglement distillation dealing with phase noise can be implemented with quantum memories whose memory processes are beam-splitter-like operations and balanced homodyne detection on the transmitted optical modes \cite{liu2018continuous}. Optical two-mode squeezed states are distributed to distant quantum nodes, during which process they suffer from phase noise. The optical states are transferred to the quantum memories at each node. A second optical two-mode squeezed state is distributed and interfered with the state of the quantum memories in the memory process, which is a beam-splitter-like operation. Conditioned on the outcome of homodyne detection on the transmitted optical mode of the beam-splitter-like operation, the entanglement is distilled. This entanglement distillation method provides highly entangled states for downstream applications under random phase fluctuations in the quantum channels used for distribution of the entangled states.

Entanglement swapping can be a Gaussian operation \cite{hoelscher2011optimal}, which mixes the two modes at the same node on a balanced beam-splitter and does homodyne detection at the output ports. The outcomes are then used at the two other nodes for the corresponding correction operation by displacing the states. Entanglement swapping can also be a non-Gaussian operation \cite{furrer2018repeaters}, in which one basically swaps the entanglement in the low-photon-number subspace. This non-Gaussian swapping protocol will of course require further Gaussification. 

As an attractive alternative to DV schemes, CV systems are also compatible with existing optical telecom systems. But in contrast to the well-developed DV quantum repeater \cite{sangouard2011quantum}, CV quantum repeaters are still in their infancy. Many existing proposals for CV quantum repeaters are similar to first-generation DV quantum repeaters, which require two-way classical communication beyond the nearest nodes \cite{dias2017quantum,furrer2018repeaters,dias2020quantum}. Some efforts have been made to develop second-generation CV repeaters which only require nearest-neighbor two-way classical communication \cite{seshadreesan2020continuous} and third-generation CV quantum repeaters which do not need two-way classical communication and are completely one-way \cite{fukui2021all}. Currently, CV quantum repeaters certainly cannot work at the repetition rates required in astronomical interferometry.  We might expect a large gain in the performance once second- and third-generation CV quantum repeaters are well established, which requires the development of CV error correction codes.



\section{Photon counting noise in the presence of squeezed states}\label{squeeze}

A famous application for squeezed states is in the interferometer used for gravitational wave detection \cite{caves1981quantum}. Squeezed states are used as a resource to reduce photon counting noise at the expense of increasing the fluctuation of radiation pressure, which is useful when the optimal laser power is not available in the practical implementation. Since our method involves squeezed states, there is a natural question: if we assume the ideal implementation of the measurement and ignore the transmission loss, which is the reason we want to use quantum teleportation instead of directly bringing the light from two telescopes to one location, can we fundamentally enhance the estimation of the coherence function over the conventional method? To the best of our understanding, the answer is unfortunately no. A short explanation is that if we consider the estimation of the coherence function using quantum estimation theory as discussed in Ref.~\cite{pearce2017optimal}, the fundamental sensitivity limit is given by the QFI. And the measurement that can saturate the optimal sensitivity for the estimation of the phase and amplitude of the coherence function can be constructed using beam-splitters and photon-number-resolved detection, respectively. The sensitivity bound given by the QFI has been optimized over all possible measurements that are physically allowed, which of course includes schemes that use squeezed states as ancilla. So, we should not expect using squeezed states to enhance the sensitivity of the astronomical interferometer. Compared with gravitational wave detection \cite{caves1981quantum}, the main difference is that in gravitational wave detection, the squeezed state is also used as an input state that is encoded with the information to measure. And in the case of the astronomical interferometer, we cannot change the state received by the astronomical source.

\begin{figure}[!tbh]
\begin{center}
\includegraphics[width=0.8\columnwidth]{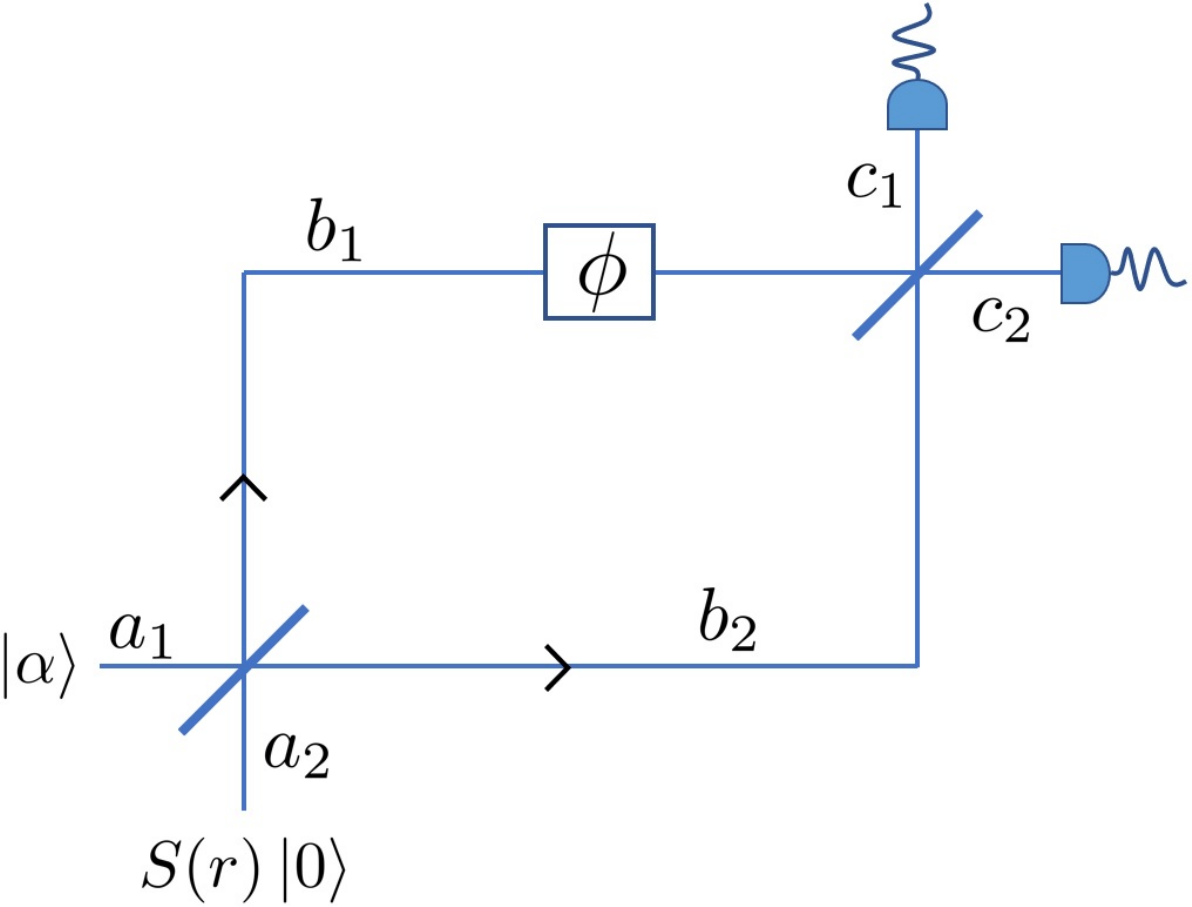}
\caption{An interferometer used for phase estimation enhanced by squeezed states. The two input ports $a_{1,2}$ are in coherent state $\ket{\alpha}$ and squeezed state $S(r)\ket{0}=\exp(\frac{r}{2}a_2^2-\frac{r}{2}{a_2^\dagger}^2)\ket{0}$. In one arm of the interferometer, an unknown phase $\phi$ is encoded on the state by unitary transformation $U(\phi)=\exp(\phi b_1^\dagger b_1)$. We measure the observable $ n=c_1^\dagger c_1-c_2^\dagger c_2$ to estimate $\phi$. }
\label{phase_estimation}
\end{center}
\end{figure}

To gain more intuition, we consider an interferometer used for phase estimation enhanced by squeezed states, which can be regarded as a simplified version of the discussion in \cite{caves1981quantum}, where we are removing all discussions related to radiation pressure. The setup is shown in Fig.~\ref{phase_estimation}. As we will see, the squeezed state in the second port of the beam-splitter can reduce the noise in the estimation when compared with the case where we leave the $a_2$ mode as the vacuum state. The unknown phase $\phi$ is estimated from the mean value of $n=c_1^\dagger c_1-c_2^\dagger c_2$ and its noise is quantified by the variance of $ n$:
\begin{equation}
\begin{aligned}
&\langle  n\rangle=(\sinh^2 r-|\alpha|^2)\cos\phi,\\
&\langle \Delta n^2\rangle=\cos^2\phi (\alpha^2+2\sinh^2r \cosh^2 r)\\
&\quad\quad\quad\quad +\sin^2\phi(\alpha^2 \cosh 2r+\sinh^2r),
\end{aligned}
\end{equation}
where $\langle.\rangle$ is the mean value of the operators for the quantum state in the output ports of the beam-splitter and we have assumed $\alpha$ is real for simplicity. Following the discussion of photon-counting error in Ref.~\cite{caves1981quantum}, we can consider $\phi$ such that $\cos\phi$ is close to zero. In this case, if we further assume the coherent state is much stronger than the squeezed state, the noise is dominated by $\sin^2\phi\alpha^2 e^{2r}/2$, which can be reduced by $r<0$. Now we consider whether this method can be applied to the astronomical interferometer. Note the squeezed state is used as the input of the interferometer. This is not possible for an astronomical interferometer since the unknown phase is only encoded in the received thermal state and the squeezed state can only be used as the ancilla of implementing the measurement.

Another intuitive question is whether we can use the squeezed LO in the implementation of the measurement. 
If we consider the implementation of heterodyne detection, in the large LO power limit, the LO noise makes no contribution to the total noise of the measurement results \cite{yuen1983noise}. So, it is not meaningful to use squeezed light as the LO because reducing the LO noise cannot reduce the  noise of the measurement, at least in any obvious way. There are some discussions that try to improve heterodyne detection using squeezed states such as Ref.~\cite{li1999sub}, which however has invited debate \cite{li2000li,ralph2000can}. So, we are satisfied with the answer that squeezed states can only be used as entanglement resources to overcome transmission loss and will not fundamentally enhance the estimation of coherence functions in an obvious way.

\section{Fisher information for heterodyne detection}\label{Appendix:heterodyne}
In this section, we derive the FI of using heterodyne detection $E(\mu,\nu)=\frac{1}{\pi^2}\ket{\mu\nu}\bra{\mu\nu}$ to estimate $|g|$ and $\theta$. The probability distribution is given by
\begin{equation}\begin{aligned}
&P(\mu,\nu)=\tr(\rho_s E(\mu,\nu))\\
&\quad\quad\quad=p \exp[\begin{pmatrix}
\mu^* & \nu^*
\end{pmatrix}B\begin{pmatrix}
\mu \\ 
\nu
\end{pmatrix}],\\
&p=\frac{1}{\pi^2}\frac{4}{4+\epsilon(4+\epsilon-\epsilon|g|^2)}\\
&B=\frac{2}{-(2+\epsilon)^2+\epsilon^2|g|^2}\begin{pmatrix}
2+\epsilon & -\epsilon g\\
-\epsilon g^* & 2+\epsilon
\end{pmatrix}.
\end{aligned}
\end{equation}
We can then calculate the FI as
\begin{equation}\begin{aligned}\label{eq:c2}
I_{\theta\theta}&=\int d^2\mu d^2\nu \frac{1}{P(\mu,\nu)}\left(\frac{\partial P(\mu,\nu)}{\partial \theta}\right)^2\\
&=\frac{2\epsilon^2|g|^2}{4+\epsilon(4+\epsilon-\epsilon|g|^2)}.
\end{aligned}\end{equation}
We can also compute the FI for estimating $|g|$ in a similar way, but deriving its analytical form is challenging.

\section{Coefficients in the derivation}\label{appendix:coefficients}
In this appendix, we provide the coefficients in the formulae of the QFI calculation in the main text from Eq.~\ref{SLDg} to Eq.~\ref{SLDtheta}.
\begin{equation}
\begin{aligned}
&a=2\epsilon^2|g|[-(1+\epsilon)(-2+\epsilon(-1+|g|^2))\\
&\quad\quad\quad\quad\quad\quad+2(2+\epsilon)y],\\
&b=\epsilon e^{-i\theta}[-\epsilon(2+\epsilon)^2+\epsilon^3|g|^4\\
&\quad\quad\quad -4(1+\epsilon)(2+\epsilon)y-4(2+\epsilon)y^2],\\
&c=2\epsilon^2|g|[-\epsilon(1+\epsilon)(-2+\epsilon(-1+|g|^2))\\
&\quad\quad+2y(2+\epsilon(4-\epsilon(-2+|g|^2)))+4(1+\epsilon)y^2],\\
&d=[\epsilon(-2+\epsilon(-1+|g|^2))-2(1+\epsilon)y]\\
&\quad\quad\times[\epsilon^2(-1+|g|^2)-4(1+y)-2\epsilon(2+y)]\\
&\quad\quad\times [\epsilon(-1+|g|^2)-2y],\\
&e=-\epsilon|g|\bigg[\frac{1}{\epsilon(-1+|g|^2)-2y}\\
&\quad\quad\quad+\frac{\epsilon}{\epsilon^2(-1+|g|^2)-4(1+y)-2\epsilon(2+y)}\bigg].
\end{aligned}
\end{equation}
\begin{equation}
\begin{aligned}
&p=\frac{i\epsilon g}{-\epsilon[-2+\epsilon(-1+|g|^2)]+2(1+\epsilon)y}.
\end{aligned}
\end{equation}

\yk{
\section{Comparison with other works about astronomical interferometer based on CV quantum teleportation}\label{appendix:comparison}

In this appendix, we highlight the differences between our work and two relevant studies on astronomical interferometry based on CV quantum teleportation: one by Zixin Huang \textit{et al.} \cite{Huang2024} and another by Bran Purvis \textit{et al.} \cite{purvis2024practical}.

In contrast to our result, Zixin Huang \textit{et al.} determined that CV teleportation-assisted telescopes offer only limited enhancement \cite{Huang2024}; however, they did not consider the use of CV repeaters to link distant telescopes. Without a CV repeater, the two-mode squeezed states used for teleportation exhibit a low effective squeezing level. As demonstrated in our protocol, with a well-performing CV repeater and a complete CV-quantum teleportation protocol where classical messages are used for adjusting the measurements, the potential benefits can be substantial.


To see how the lack of a CV repeater works out in practice, let us imagine distributing an entangled state through a lossy channel and directly using this noisy state for teleportation, as done in \cite{Huang2024}. In this scenario, the noise from the lossy channel, along with additional noise introduced by imperfections in the subsequent teleportation operations or insufficient squeezing level of the two-mode squeezed state, will be transferred to the teleported states. In other words, the noise from the lossy channel is not effectively mitigated by the teleportation process, as the entangled state is highly susceptible to noise during transmission through the noisy channel, and the teleportation process itself introduces a significant amount of additional noise.
As a result, the teleported state is likely to be even more noisy than a state obtained by directly transmitting the stellar light through the lossy channel without the added complexity of entanglement generation, distribution, and teleportation. Indeed,  Huang \textit{et al.} found that the advantage of using teleportation with such an entangled state in a repeaterless setting is very limited \cite{Huang2024}, which aligns with this reasoning.
While the investigation of a repeaterless scenario is reasonable and interesting for near-term applications, the deployment of quantum repeaters will ultimately be necessary to realize the full potential of quantum-enhanced astronomical interferometry as in Ref. \cite{gottesman2012longer,khabiboulline2019optical,khabiboulline2019quantum}. 

Huang \textit{et al.} observed that there are certain parameter regimes in which CV teleportation can outperform direct detection and local heterodyne approaches \cite{Huang2024}. However, their analysis relies on directly utilizing noisy entangled states without an entanglement distillation step. In contrast, our protocol employs CV teleportation using entangled states that can be distilled through a CV quantum repeater, thereby mitigating transmission loss. As long as such a repeater is sufficiently efficient, our scheme consistently outperforms direct detection. And our scheme surpasses local heterodyne detection whenever the source intensity is not too high, as shown in Fig.~\ref{compare}.

We now highlight the key differences between our work and that of Bran Purvis \textit{et al.}~\cite{purvis2024practical}. 
Although their protocol also utilizes entanglement resources in the form of two-mode squeezed states shared between distant telescopes, it relies solely on Gaussian measurements, in contrast to our scheme, which employs non-Gaussian photon-number-resolving detection following quantum teleportation.
As a result, a major drawback of their approach is that the performance, quantified by the Fisher information (FI) $F$, scales as $F \propto \epsilon^2$. In contrast, the advantage of an astronomical interferometer assisted by a quantum repeater is that its FI scales as $F \propto \epsilon$, as in our protocol when $\epsilon \ll 1$, as demonstrated in Eq.~\ref{FI}. Note that the initial proposal for an astronomical interferometer based on DV quantum repeaters by Daniel Gottesman \textit{et al.} \cite{gottesman2012longer} also achieves the FI scaling $F\propto \epsilon$.
When $\epsilon \ll 1$, this difference in scaling leads to a substantial performance gap between the protocol in \cite{purvis2024practical} and ours. 

In fact, simple local heterodyne detection alone yields a Fisher information scaling as $F \propto \epsilon^2$, as shown in Eq.~\ref{eq:c2}. Since the FI in the work of Bran Purvis \textit{et al.}~\cite{purvis2024practical} also follows the same scaling, $F \propto \epsilon^2$, its performance is comparable to the case without the complex steps of entanglement distribution and teleportation. As noted in Ref.~\cite{tsang2011fundamental}, the scaling behavior of the FI with respect to $\epsilon$ is a key characteristic of quantum-enhanced astronomical interferometry, which relies on shared entanglement.
Given that the approach proposed in Ref.~\cite{purvis2024practical} also relies on shared entanglement but achieves only $F \propto \epsilon^2$, it does not provide an efficient method to fully exploit the shared entanglement between distant telescopes.

}

\vspace{1cm}

\bibliography{CV_teleportation}

\begin{thebibliography}{78}%
\makeatletter
\providecommand \@ifxundefined [1]{%
 \@ifx{#1\undefined}
}%
\providecommand \@ifnum [1]{%
 \ifnum #1\expandafter \@firstoftwo
 \else \expandafter \@secondoftwo
 \fi
}%
\providecommand \@ifx [1]{%
 \ifx #1\expandafter \@firstoftwo
 \else \expandafter \@secondoftwo
 \fi
}%
\providecommand \natexlab [1]{#1}%
\providecommand \enquote  [1]{``#1''}%
\providecommand \bibnamefont  [1]{#1}%
\providecommand \bibfnamefont [1]{#1}%
\providecommand \citenamefont [1]{#1}%
\providecommand \href@noop [0]{\@secondoftwo}%
\providecommand \href [0]{\begingroup \@sanitize@url \@href}%
\providecommand \@href[1]{\@@startlink{#1}\@@href}%
\providecommand \@@href[1]{\endgroup#1\@@endlink}%
\providecommand \@sanitize@url [0]{\catcode `\\12\catcode `\$12\catcode
  `\&12\catcode `\#12\catcode `\^12\catcode `\_12\catcode `\%12\relax}%
\providecommand \@@startlink[1]{}%
\providecommand \@@endlink[0]{}%
\providecommand \url  [0]{\begingroup\@sanitize@url \@url }%
\providecommand \@url [1]{\endgroup\@href {#1}{\urlprefix }}%
\providecommand \urlprefix  [0]{URL }%
\providecommand \Eprint [0]{\href }%
\providecommand \doibase [0]{https://doi.org/}%
\providecommand \selectlanguage [0]{\@gobble}%
\providecommand \bibinfo  [0]{\@secondoftwo}%
\providecommand \bibfield  [0]{\@secondoftwo}%
\providecommand \translation [1]{[#1]}%
\providecommand \BibitemOpen [0]{}%
\providecommand \bibitemStop [0]{}%
\providecommand \bibitemNoStop [0]{.\EOS\space}%
\providecommand \EOS [0]{\spacefactor3000\relax}%
\providecommand \BibitemShut  [1]{\csname bibitem#1\endcsname}%
\let\auto@bib@innerbib\@empty
\bibitem [{\citenamefont {Zernike}(1938)}]{zernike1938concept}%
  \BibitemOpen
  \bibfield  {author} {\bibinfo {author} {\bibfnamefont {F.}~\bibnamefont
  {Zernike}},\ }\bibfield  {title} {\bibinfo {title} {The concept of degree of
  coherence and its application to optical problems},\ }\href@noop {}
  {\bibfield  {journal} {\bibinfo  {journal} {Physica}\ }\textbf {\bibinfo
  {volume} {5}},\ \bibinfo {pages} {785} (\bibinfo {year} {1938})}\BibitemShut
  {NoStop}%
\bibitem [{\citenamefont {Collaboration}\ \emph
  {et~al.}(2019{\natexlab{a}})\citenamefont {Collaboration}, \citenamefont
  {Akiyama}, \citenamefont {Alberdi}, \citenamefont {Alef}, \citenamefont
  {Asada}, \citenamefont {AZULY} \emph {et~al.}}]{collaboration2019first}%
  \BibitemOpen
  \bibfield  {author} {\bibinfo {author} {\bibfnamefont {E.~H.~T.}\
  \bibnamefont {Collaboration}}, \bibinfo {author} {\bibfnamefont
  {K.}~\bibnamefont {Akiyama}}, \bibinfo {author} {\bibfnamefont
  {A.}~\bibnamefont {Alberdi}}, \bibinfo {author} {\bibfnamefont
  {W.}~\bibnamefont {Alef}}, \bibinfo {author} {\bibfnamefont {K.}~\bibnamefont
  {Asada}}, \bibinfo {author} {\bibfnamefont {R.}~\bibnamefont {AZULY}}, \emph
  {et~al.},\ }\bibfield  {title} {\bibinfo {title} {First m87 event horizon
  telescope results. i. the shadow of the supermassive black hole},\
  }\href@noop {} {\bibfield  {journal} {\bibinfo  {journal} {Astrophys. J.
  Lett}\ }\textbf {\bibinfo {volume} {875}},\ \bibinfo {pages} {L1} (\bibinfo
  {year} {2019}{\natexlab{a}})}\BibitemShut {NoStop}%
\bibitem [{\citenamefont {Wilson}\ \emph {et~al.}(2009)\citenamefont {Wilson},
  \citenamefont {Rohlfs},\ and\ \citenamefont
  {H{\"u}ttemeister}}]{wilson2009tools}%
  \BibitemOpen
  \bibfield  {author} {\bibinfo {author} {\bibfnamefont {T.~L.}\ \bibnamefont
  {Wilson}}, \bibinfo {author} {\bibfnamefont {K.}~\bibnamefont {Rohlfs}},\
  and\ \bibinfo {author} {\bibfnamefont {S.}~\bibnamefont {H{\"u}ttemeister}},\
  }\href@noop {} {\emph {\bibinfo {title} {Tools of radio astronomy}}},\
  Vol.~\bibinfo {volume} {5}\ (\bibinfo  {publisher} {Springer},\ \bibinfo
  {year} {2009})\BibitemShut {NoStop}%
\bibitem [{\citenamefont {Monnier}(2003)}]{monnier2003optical}%
  \BibitemOpen
  \bibfield  {author} {\bibinfo {author} {\bibfnamefont {J.~D.}\ \bibnamefont
  {Monnier}},\ }\bibfield  {title} {\bibinfo {title} {Optical interferometry in
  astronomy},\ }\href@noop {} {\bibfield  {journal} {\bibinfo  {journal}
  {Reports on Progress in Physics}\ }\textbf {\bibinfo {volume} {66}},\
  \bibinfo {pages} {789} (\bibinfo {year} {2003})}\BibitemShut {NoStop}%
\bibitem [{\citenamefont {Tsang}(2011)}]{tsang2011quantum}%
  \BibitemOpen
  \bibfield  {author} {\bibinfo {author} {\bibfnamefont {M.}~\bibnamefont
  {Tsang}},\ }\bibfield  {title} {\bibinfo {title} {Quantum nonlocality in
  weak-thermal-light interferometry},\ }\href@noop {} {\bibfield  {journal}
  {\bibinfo  {journal} {Phys. Rev. Lett.}\ }\textbf {\bibinfo {volume} {107}},\
  \bibinfo {pages} {270402} (\bibinfo {year} {2011})}\BibitemShut {NoStop}%
\bibitem [{\citenamefont {Gottesman}\ \emph {et~al.}(2012)\citenamefont
  {Gottesman}, \citenamefont {Jennewein},\ and\ \citenamefont
  {Croke}}]{gottesman2012longer}%
  \BibitemOpen
  \bibfield  {author} {\bibinfo {author} {\bibfnamefont {D.}~\bibnamefont
  {Gottesman}}, \bibinfo {author} {\bibfnamefont {T.}~\bibnamefont
  {Jennewein}},\ and\ \bibinfo {author} {\bibfnamefont {S.}~\bibnamefont
  {Croke}},\ }\bibfield  {title} {\bibinfo {title} {Longer-baseline telescopes
  using quantum repeaters},\ }\href@noop {} {\bibfield  {journal} {\bibinfo
  {journal} {Phys. Rev. Lett.}\ }\textbf {\bibinfo {volume} {109}},\ \bibinfo
  {pages} {070503} (\bibinfo {year} {2012})}\BibitemShut {NoStop}%
\bibitem [{\citenamefont {Khabiboulline}\ \emph
  {et~al.}(2019{\natexlab{a}})\citenamefont {Khabiboulline}, \citenamefont
  {Borregaard}, \citenamefont {De~Greve},\ and\ \citenamefont
  {Lukin}}]{khabiboulline2019optical}%
  \BibitemOpen
  \bibfield  {author} {\bibinfo {author} {\bibfnamefont {E.~T.}\ \bibnamefont
  {Khabiboulline}}, \bibinfo {author} {\bibfnamefont {J.}~\bibnamefont
  {Borregaard}}, \bibinfo {author} {\bibfnamefont {K.}~\bibnamefont
  {De~Greve}},\ and\ \bibinfo {author} {\bibfnamefont {M.~D.}\ \bibnamefont
  {Lukin}},\ }\bibfield  {title} {\bibinfo {title} {Optical interferometry with
  quantum networks},\ }\href@noop {} {\bibfield  {journal} {\bibinfo  {journal}
  {Phys. Rev. Lett.}\ }\textbf {\bibinfo {volume} {123}},\ \bibinfo {pages}
  {070504} (\bibinfo {year} {2019}{\natexlab{a}})}\BibitemShut {NoStop}%
\bibitem [{\citenamefont {Khabiboulline}\ \emph
  {et~al.}(2019{\natexlab{b}})\citenamefont {Khabiboulline}, \citenamefont
  {Borregaard}, \citenamefont {De~Greve},\ and\ \citenamefont
  {Lukin}}]{khabiboulline2019quantum}%
  \BibitemOpen
  \bibfield  {author} {\bibinfo {author} {\bibfnamefont {E.~T.}\ \bibnamefont
  {Khabiboulline}}, \bibinfo {author} {\bibfnamefont {J.}~\bibnamefont
  {Borregaard}}, \bibinfo {author} {\bibfnamefont {K.}~\bibnamefont
  {De~Greve}},\ and\ \bibinfo {author} {\bibfnamefont {M.~D.}\ \bibnamefont
  {Lukin}},\ }\bibfield  {title} {\bibinfo {title} {Quantum-assisted telescope
  arrays},\ }\href@noop {} {\bibfield  {journal} {\bibinfo  {journal} {Phys.
  Rev. A}\ }\textbf {\bibinfo {volume} {100}},\ \bibinfo {pages} {022316}
  (\bibinfo {year} {2019}{\natexlab{b}})}\BibitemShut {NoStop}%
\bibitem [{\citenamefont {Huang}\ \emph {et~al.}(2022)\citenamefont {Huang},
  \citenamefont {Brennen},\ and\ \citenamefont {Ouyang}}]{Huang2022}%
  \BibitemOpen
  \bibfield  {author} {\bibinfo {author} {\bibfnamefont {Z.}~\bibnamefont
  {Huang}}, \bibinfo {author} {\bibfnamefont {G.~K.}\ \bibnamefont {Brennen}},\
  and\ \bibinfo {author} {\bibfnamefont {Y.}~\bibnamefont {Ouyang}},\
  }\bibfield  {title} {\bibinfo {title} {Imaging stars with quantum error
  correction},\ }\href {https://doi.org/10.1103/PhysRevLett.129.210502}
  {\bibfield  {journal} {\bibinfo  {journal} {Phys. Rev. Lett.}\ }\textbf
  {\bibinfo {volume} {129}},\ \bibinfo {pages} {210502} (\bibinfo {year}
  {2022})}\BibitemShut {NoStop}%
\bibitem [{\citenamefont {Czupryniak}\ \emph {et~al.}(2023)\citenamefont
  {Czupryniak}, \citenamefont {Steinmetz}, \citenamefont {Kwiat},\ and\
  \citenamefont {Jordan}}]{Czupryniak2023}%
  \BibitemOpen
  \bibfield  {author} {\bibinfo {author} {\bibfnamefont {R.}~\bibnamefont
  {Czupryniak}}, \bibinfo {author} {\bibfnamefont {J.}~\bibnamefont
  {Steinmetz}}, \bibinfo {author} {\bibfnamefont {P.~G.}\ \bibnamefont
  {Kwiat}},\ and\ \bibinfo {author} {\bibfnamefont {A.~N.}\ \bibnamefont
  {Jordan}},\ }\bibfield  {title} {\bibinfo {title} {Optimal qubit circuits for
  quantum-enhanced telescopes},\ }\href
  {https://doi.org/10.1103/PhysRevA.108.052408} {\bibfield  {journal} {\bibinfo
   {journal} {Phys. Rev. A}\ }\textbf {\bibinfo {volume} {108}},\ \bibinfo
  {pages} {052408} (\bibinfo {year} {2023})}\BibitemShut {NoStop}%
\bibitem [{\citenamefont {Sangouard}\ \emph {et~al.}(2011)\citenamefont
  {Sangouard}, \citenamefont {Simon}, \citenamefont {De~Riedmatten},\ and\
  \citenamefont {Gisin}}]{sangouard2011quantum}%
  \BibitemOpen
  \bibfield  {author} {\bibinfo {author} {\bibfnamefont {N.}~\bibnamefont
  {Sangouard}}, \bibinfo {author} {\bibfnamefont {C.}~\bibnamefont {Simon}},
  \bibinfo {author} {\bibfnamefont {H.}~\bibnamefont {De~Riedmatten}},\ and\
  \bibinfo {author} {\bibfnamefont {N.}~\bibnamefont {Gisin}},\ }\bibfield
  {title} {\bibinfo {title} {Quantum repeaters based on atomic ensembles and
  linear optics},\ }\href@noop {} {\bibfield  {journal} {\bibinfo  {journal}
  {Rev. Mod. Phys.}\ }\textbf {\bibinfo {volume} {83}},\ \bibinfo {pages} {33}
  (\bibinfo {year} {2011})}\BibitemShut {NoStop}%
\bibitem [{\citenamefont {Czupryniak}\ \emph {et~al.}(2022)\citenamefont
  {Czupryniak}, \citenamefont {Chitambar}, \citenamefont {Steinmetz},\ and\
  \citenamefont {Jordan}}]{Czupryniak2022}%
  \BibitemOpen
  \bibfield  {author} {\bibinfo {author} {\bibfnamefont {R.}~\bibnamefont
  {Czupryniak}}, \bibinfo {author} {\bibfnamefont {E.}~\bibnamefont
  {Chitambar}}, \bibinfo {author} {\bibfnamefont {J.}~\bibnamefont
  {Steinmetz}},\ and\ \bibinfo {author} {\bibfnamefont {A.~N.}\ \bibnamefont
  {Jordan}},\ }\bibfield  {title} {\bibinfo {title} {Quantum telescopy clock
  games},\ }\href {https://doi.org/10.1103/PhysRevA.106.032424} {\bibfield
  {journal} {\bibinfo  {journal} {Phys. Rev. A}\ }\textbf {\bibinfo {volume}
  {106}},\ \bibinfo {pages} {032424} (\bibinfo {year} {2022})}\BibitemShut
  {NoStop}%
\bibitem [{\citenamefont {Marchese}\ and\ \citenamefont
  {Kok}(2023)}]{Marchese2023}%
  \BibitemOpen
  \bibfield  {author} {\bibinfo {author} {\bibfnamefont {M.~M.}\ \bibnamefont
  {Marchese}}\ and\ \bibinfo {author} {\bibfnamefont {P.}~\bibnamefont {Kok}},\
  }\bibfield  {title} {\bibinfo {title} {Large baseline optical imaging
  assisted by single photons and linear quantum optics},\ }\href
  {https://doi.org/10.1103/PhysRevLett.130.160801} {\bibfield  {journal}
  {\bibinfo  {journal} {Phys. Rev. Lett.}\ }\textbf {\bibinfo {volume} {130}},\
  \bibinfo {pages} {160801} (\bibinfo {year} {2023})}\BibitemShut {NoStop}%
\bibitem [{\citenamefont {Zhang}\ and\ \citenamefont
  {Jennewein}(2025)}]{zhang2025}%
  \BibitemOpen
  \bibfield  {author} {\bibinfo {author} {\bibfnamefont {Y.}~\bibnamefont
  {Zhang}}\ and\ \bibinfo {author} {\bibfnamefont {T.}~\bibnamefont
  {Jennewein}},\ }\href {https://arxiv.org/abs/2501.16670} {\bibinfo {title}
  {Criteria for optimal entanglement-assisted long baseline imaging protocols}}
  (\bibinfo {year} {2025}),\ \Eprint {https://arxiv.org/abs/2501.16670}
  {arXiv:2501.16670 [quant-ph]} \BibitemShut {NoStop}%
\bibitem [{\citenamefont {Modak}\ and\ \citenamefont {Kok}(2024)}]{modak2024}%
  \BibitemOpen
  \bibfield  {author} {\bibinfo {author} {\bibfnamefont {S.}~\bibnamefont
  {Modak}}\ and\ \bibinfo {author} {\bibfnamefont {P.}~\bibnamefont {Kok}},\
  }\href {https://arxiv.org/abs/2412.16571} {\bibinfo {title} {Large baseline
  quantum telescopes assisted by partially distinguishable photons}} (\bibinfo
  {year} {2024}),\ \Eprint {https://arxiv.org/abs/2412.16571} {arXiv:2412.16571
  [quant-ph]} \BibitemShut {NoStop}%
\bibitem [{\citenamefont {Vaidman}(1994)}]{vaidman1994teleportation}%
  \BibitemOpen
  \bibfield  {author} {\bibinfo {author} {\bibfnamefont {L.}~\bibnamefont
  {Vaidman}},\ }\bibfield  {title} {\bibinfo {title} {Teleportation of quantum
  states},\ }\href@noop {} {\bibfield  {journal} {\bibinfo  {journal} {Phys.
  Rev. A}\ }\textbf {\bibinfo {volume} {49}},\ \bibinfo {pages} {1473}
  (\bibinfo {year} {1994})}\BibitemShut {NoStop}%
\bibitem [{\citenamefont {Braunstein}\ and\ \citenamefont
  {Kimble}(1998)}]{braunstein1998teleportation}%
  \BibitemOpen
  \bibfield  {author} {\bibinfo {author} {\bibfnamefont {S.~L.}\ \bibnamefont
  {Braunstein}}\ and\ \bibinfo {author} {\bibfnamefont {H.~J.}\ \bibnamefont
  {Kimble}},\ }\bibfield  {title} {\bibinfo {title} {Teleportation of
  continuous quantum variables},\ }\href@noop {} {\bibfield  {journal}
  {\bibinfo  {journal} {Phys. Rev. Lett.}\ }\textbf {\bibinfo {volume} {80}},\
  \bibinfo {pages} {869} (\bibinfo {year} {1998})}\BibitemShut {NoStop}%
\bibitem [{\citenamefont {Pirandola}\ and\ \citenamefont
  {Mancini}(2006)}]{pirandola2006quantum}%
  \BibitemOpen
  \bibfield  {author} {\bibinfo {author} {\bibfnamefont {S.}~\bibnamefont
  {Pirandola}}\ and\ \bibinfo {author} {\bibfnamefont {S.}~\bibnamefont
  {Mancini}},\ }\bibfield  {title} {\bibinfo {title} {Quantum teleportation
  with continuous variables: A survey},\ }\href@noop {} {\bibfield  {journal}
  {\bibinfo  {journal} {Laser Physics}\ }\textbf {\bibinfo {volume} {16}},\
  \bibinfo {pages} {1418} (\bibinfo {year} {2006})}\BibitemShut {NoStop}%
\bibitem [{\citenamefont {Weedbrook}\ \emph {et~al.}(2012)\citenamefont
  {Weedbrook}, \citenamefont {Pirandola}, \citenamefont
  {Garc{\'\i}a-Patr{\'o}n}, \citenamefont {Cerf}, \citenamefont {Ralph},
  \citenamefont {Shapiro},\ and\ \citenamefont
  {Lloyd}}]{weedbrook2012gaussian}%
  \BibitemOpen
  \bibfield  {author} {\bibinfo {author} {\bibfnamefont {C.}~\bibnamefont
  {Weedbrook}}, \bibinfo {author} {\bibfnamefont {S.}~\bibnamefont
  {Pirandola}}, \bibinfo {author} {\bibfnamefont {R.}~\bibnamefont
  {Garc{\'\i}a-Patr{\'o}n}}, \bibinfo {author} {\bibfnamefont {N.~J.}\
  \bibnamefont {Cerf}}, \bibinfo {author} {\bibfnamefont {T.~C.}\ \bibnamefont
  {Ralph}}, \bibinfo {author} {\bibfnamefont {J.~H.}\ \bibnamefont {Shapiro}},\
  and\ \bibinfo {author} {\bibfnamefont {S.}~\bibnamefont {Lloyd}},\ }\bibfield
   {title} {\bibinfo {title} {Gaussian quantum information},\ }\href@noop {}
  {\bibfield  {journal} {\bibinfo  {journal} {Rev. Mod. Phys.}\ }\textbf
  {\bibinfo {volume} {84}},\ \bibinfo {pages} {621} (\bibinfo {year}
  {2012})}\BibitemShut {NoStop}%
\bibitem [{\citenamefont {Dong}\ \emph {et~al.}(2008)\citenamefont {Dong},
  \citenamefont {Heersink}, \citenamefont {Corney}, \citenamefont {Drummond},
  \citenamefont {Andersen},\ and\ \citenamefont
  {Leuchs}}]{dong2008experimental}%
  \BibitemOpen
  \bibfield  {author} {\bibinfo {author} {\bibfnamefont {R.}~\bibnamefont
  {Dong}}, \bibinfo {author} {\bibfnamefont {J.}~\bibnamefont {Heersink}},
  \bibinfo {author} {\bibfnamefont {J.~F.}\ \bibnamefont {Corney}}, \bibinfo
  {author} {\bibfnamefont {P.~D.}\ \bibnamefont {Drummond}}, \bibinfo {author}
  {\bibfnamefont {U.~L.}\ \bibnamefont {Andersen}},\ and\ \bibinfo {author}
  {\bibfnamefont {G.}~\bibnamefont {Leuchs}},\ }\bibfield  {title} {\bibinfo
  {title} {Experimental evidence for raman-induced limits to efficient
  squeezing in optical fibers},\ }\href@noop {} {\bibfield  {journal} {\bibinfo
   {journal} {Optics letters}\ }\textbf {\bibinfo {volume} {33}},\ \bibinfo
  {pages} {116} (\bibinfo {year} {2008})}\BibitemShut {NoStop}%
\bibitem [{\citenamefont {Mandel}\ and\ \citenamefont
  {Wolf}(1995)}]{mandel1995optical}%
  \BibitemOpen
  \bibfield  {author} {\bibinfo {author} {\bibfnamefont {L.}~\bibnamefont
  {Mandel}}\ and\ \bibinfo {author} {\bibfnamefont {E.}~\bibnamefont {Wolf}},\
  }\href@noop {} {\emph {\bibinfo {title} {Optical coherence and quantum
  optics}}}\ (\bibinfo  {publisher} {Cambridge university press},\ \bibinfo
  {year} {1995})\BibitemShut {NoStop}%
\bibitem [{\citenamefont {Duan}\ and\ \citenamefont
  {Guo}(1997)}]{duan1997influence}%
  \BibitemOpen
  \bibfield  {author} {\bibinfo {author} {\bibfnamefont {L.-M.}\ \bibnamefont
  {Duan}}\ and\ \bibinfo {author} {\bibfnamefont {G.-C.}\ \bibnamefont {Guo}},\
  }\bibfield  {title} {\bibinfo {title} {Influence of noise on the fidelity and
  the entanglement fidelity of states},\ }\href@noop {} {\bibfield  {journal}
  {\bibinfo  {journal} {Quantum and Semiclassical Optics: Journal of the
  European Optical Society Part B}\ }\textbf {\bibinfo {volume} {9}},\ \bibinfo
  {pages} {953} (\bibinfo {year} {1997})}\BibitemShut {NoStop}%
\bibitem [{\citenamefont {Chizhov}\ \emph {et~al.}(2001)\citenamefont
  {Chizhov}, \citenamefont {Schmidt}, \citenamefont {Kn{\"o}ll},\ and\
  \citenamefont {Welsch}}]{chizhov2001propagation}%
  \BibitemOpen
  \bibfield  {author} {\bibinfo {author} {\bibfnamefont {A.}~\bibnamefont
  {Chizhov}}, \bibinfo {author} {\bibfnamefont {E.}~\bibnamefont {Schmidt}},
  \bibinfo {author} {\bibfnamefont {L.}~\bibnamefont {Kn{\"o}ll}},\ and\
  \bibinfo {author} {\bibfnamefont {D.}~\bibnamefont {Welsch}},\ }\bibfield
  {title} {\bibinfo {title} {Propagation of entangled light pulses through
  dispersing and absorbing channels},\ }\href@noop {} {\bibfield  {journal}
  {\bibinfo  {journal} {Journal of Optics B: Quantum and Semiclassical Optics}\
  }\textbf {\bibinfo {volume} {3}},\ \bibinfo {pages} {77} (\bibinfo {year}
  {2001})}\BibitemShut {NoStop}%
\bibitem [{\citenamefont {Scheel}\ \emph {et~al.}(2001)\citenamefont {Scheel},
  \citenamefont {Opatrny},\ and\ \citenamefont
  {Welsch}}]{scheel2001entanglement}%
  \BibitemOpen
  \bibfield  {author} {\bibinfo {author} {\bibfnamefont {S.}~\bibnamefont
  {Scheel}}, \bibinfo {author} {\bibfnamefont {T.}~\bibnamefont {Opatrny}},\
  and\ \bibinfo {author} {\bibfnamefont {D.-G.}\ \bibnamefont {Welsch}},\
  }\bibfield  {title} {\bibinfo {title} {Entanglement degradation of a two-mode
  squeezed vacuum in absorbing and amplifying optical fibers},\ }\href@noop {}
  {\bibfield  {journal} {\bibinfo  {journal} {Optics and Spectroscopy}\
  }\textbf {\bibinfo {volume} {91}},\ \bibinfo {pages} {411} (\bibinfo {year}
  {2001})}\BibitemShut {NoStop}%
\bibitem [{\citenamefont {Chizhov}\ \emph {et~al.}(2002)\citenamefont
  {Chizhov}, \citenamefont {Kn{\"o}ll},\ and\ \citenamefont
  {Welsch}}]{chizhov2002continuous}%
  \BibitemOpen
  \bibfield  {author} {\bibinfo {author} {\bibfnamefont {A.}~\bibnamefont
  {Chizhov}}, \bibinfo {author} {\bibfnamefont {L.}~\bibnamefont {Kn{\"o}ll}},\
  and\ \bibinfo {author} {\bibfnamefont {D.-G.}\ \bibnamefont {Welsch}},\
  }\bibfield  {title} {\bibinfo {title} {Continuous-variable quantum
  teleportation through lossy channels},\ }\href@noop {} {\bibfield  {journal}
  {\bibinfo  {journal} {Phys. Rev. A}\ }\textbf {\bibinfo {volume} {65}},\
  \bibinfo {pages} {022310} (\bibinfo {year} {2002})}\BibitemShut {NoStop}%
\bibitem [{\citenamefont {Ban}\ \emph {et~al.}(2002)\citenamefont {Ban},
  \citenamefont {Sasaki},\ and\ \citenamefont {Takeoka}}]{ban2002continuous}%
  \BibitemOpen
  \bibfield  {author} {\bibinfo {author} {\bibfnamefont {M.}~\bibnamefont
  {Ban}}, \bibinfo {author} {\bibfnamefont {M.}~\bibnamefont {Sasaki}},\ and\
  \bibinfo {author} {\bibfnamefont {M.}~\bibnamefont {Takeoka}},\ }\bibfield
  {title} {\bibinfo {title} {Continuous variable teleportation as a generalized
  thermalizing quantum channel},\ }\href@noop {} {\bibfield  {journal}
  {\bibinfo  {journal} {Journal of Physics A: Mathematical and General}\
  }\textbf {\bibinfo {volume} {35}},\ \bibinfo {pages} {L401} (\bibinfo {year}
  {2002})}\BibitemShut {NoStop}%
\bibitem [{\citenamefont {Helstrom}(1976)}]{helstrom1976quantum}%
  \BibitemOpen
  \bibfield  {author} {\bibinfo {author} {\bibfnamefont {C.}~\bibnamefont
  {Helstrom}},\ }\bibfield  {title} {\bibinfo {title} {Quantum detection and
  estimation theory, academic press},\ }\href@noop {} {\bibfield  {journal}
  {\bibinfo  {journal} {New York}\ } (\bibinfo {year} {1976})}\BibitemShut
  {NoStop}%
\bibitem [{\citenamefont {Monras}(2013)}]{monras2013phase}%
  \BibitemOpen
  \bibfield  {author} {\bibinfo {author} {\bibfnamefont {A.}~\bibnamefont
  {Monras}},\ }\bibfield  {title} {\bibinfo {title} {Phase space formalism for
  quantum estimation of gaussian states},\ }\href@noop {} {\bibfield  {journal}
  {\bibinfo  {journal} {arXiv preprint arXiv:1303.3682}\ } (\bibinfo {year}
  {2013})}\BibitemShut {NoStop}%
\bibitem [{\citenamefont {Gao}\ and\ \citenamefont
  {Lee}(2014)}]{gao2014bounds}%
  \BibitemOpen
  \bibfield  {author} {\bibinfo {author} {\bibfnamefont {Y.}~\bibnamefont
  {Gao}}\ and\ \bibinfo {author} {\bibfnamefont {H.}~\bibnamefont {Lee}},\
  }\bibfield  {title} {\bibinfo {title} {Bounds on quantum multiple-parameter
  estimation with gaussian state},\ }\href@noop {} {\bibfield  {journal}
  {\bibinfo  {journal} {The European Physical Journal D}\ }\textbf {\bibinfo
  {volume} {68}},\ \bibinfo {pages} {1} (\bibinfo {year} {2014})}\BibitemShut
  {NoStop}%
\bibitem [{\citenamefont {Huang}\ \emph {et~al.}(2024)\citenamefont {Huang},
  \citenamefont {Baragiola}, \citenamefont {Menicucci},\ and\ \citenamefont
  {Wilde}}]{Huang2024}%
  \BibitemOpen
  \bibfield  {author} {\bibinfo {author} {\bibfnamefont {Z.}~\bibnamefont
  {Huang}}, \bibinfo {author} {\bibfnamefont {B.~Q.}\ \bibnamefont
  {Baragiola}}, \bibinfo {author} {\bibfnamefont {N.~C.}\ \bibnamefont
  {Menicucci}},\ and\ \bibinfo {author} {\bibfnamefont {M.~M.}\ \bibnamefont
  {Wilde}},\ }\href@noop {} {\bibinfo {title} {Limited quantum advantage for
  stellar interferometry via continuous-variable teleportation}} (\bibinfo
  {year} {2024}),\ \Eprint {https://arxiv.org/abs/2311.05159} {arXiv:2311.05159
  [quant-ph]} \BibitemShut {NoStop}%
\bibitem [{\citenamefont {Nielsen}\ and\ \citenamefont
  {Chuang}(2002)}]{nielsen2002quantum}%
  \BibitemOpen
  \bibfield  {author} {\bibinfo {author} {\bibfnamefont {M.~A.}\ \bibnamefont
  {Nielsen}}\ and\ \bibinfo {author} {\bibfnamefont {I.}~\bibnamefont
  {Chuang}},\ }\href@noop {} {\bibinfo {title} {Quantum computation and quantum
  information}} (\bibinfo {year} {2002})\BibitemShut {NoStop}%
\bibitem [{\citenamefont {Braunstein}\ and\ \citenamefont
  {Caves}(1994)}]{braunstein1994statistical}%
  \BibitemOpen
  \bibfield  {author} {\bibinfo {author} {\bibfnamefont {S.~L.}\ \bibnamefont
  {Braunstein}}\ and\ \bibinfo {author} {\bibfnamefont {C.~M.}\ \bibnamefont
  {Caves}},\ }\bibfield  {title} {\bibinfo {title} {Statistical distance and
  the geometry of quantum states},\ }\href@noop {} {\bibfield  {journal}
  {\bibinfo  {journal} {Phys. Rev. Lett.}\ }\textbf {\bibinfo {volume} {72}},\
  \bibinfo {pages} {3439} (\bibinfo {year} {1994})}\BibitemShut {NoStop}%
\bibitem [{\citenamefont {Paris}(2009)}]{paris2009quantum}%
  \BibitemOpen
  \bibfield  {author} {\bibinfo {author} {\bibfnamefont {M.~G.}\ \bibnamefont
  {Paris}},\ }\bibfield  {title} {\bibinfo {title} {Quantum estimation for
  quantum technology},\ }\href@noop {} {\bibfield  {journal} {\bibinfo
  {journal} {International Journal of Quantum Information}\ }\textbf {\bibinfo
  {volume} {7}},\ \bibinfo {pages} {125} (\bibinfo {year} {2009})}\BibitemShut
  {NoStop}%
\bibitem [{\citenamefont {Collaboration}\ \emph
  {et~al.}(2019{\natexlab{b}})\citenamefont {Collaboration}, \citenamefont
  {Akiyama}, \citenamefont {Alberdi}, \citenamefont {Alef}, \citenamefont
  {Asada}, \citenamefont {Azuly} \emph {et~al.}}]{EHT2019}%
  \BibitemOpen
  \bibfield  {author} {\bibinfo {author} {\bibfnamefont {E.~H.~T.}\
  \bibnamefont {Collaboration}}, \bibinfo {author} {\bibfnamefont
  {K.}~\bibnamefont {Akiyama}}, \bibinfo {author} {\bibfnamefont
  {A.}~\bibnamefont {Alberdi}}, \bibinfo {author} {\bibfnamefont
  {W.}~\bibnamefont {Alef}}, \bibinfo {author} {\bibfnamefont {K.}~\bibnamefont
  {Asada}}, \bibinfo {author} {\bibfnamefont {R.}~\bibnamefont {Azuly}}, \emph
  {et~al.},\ }\bibfield  {title} {\bibinfo {title} {First m87 event horizon
  telescope results. i. the shadow of the supermassive black hole},\
  }\href@noop {} {\bibfield  {journal} {\bibinfo  {journal} {Astrophys. J.
  Lett}\ }\textbf {\bibinfo {volume} {875}},\ \bibinfo {pages} {L1} (\bibinfo
  {year} {2019}{\natexlab{b}})}\BibitemShut {NoStop}%
\bibitem [{\citenamefont {Boyajian}\ \emph {et~al.}(2015)\citenamefont
  {Boyajian}, \citenamefont {von Braun}, \citenamefont {Feiden}, \citenamefont
  {Huber}, \citenamefont {Basu}, \citenamefont {Demarque}, \citenamefont
  {Fischer}, \citenamefont {Schaefer}, \citenamefont {Mann}, \citenamefont
  {White} \emph {et~al.}}]{boyajian2015stellar}%
  \BibitemOpen
  \bibfield  {author} {\bibinfo {author} {\bibfnamefont {T.}~\bibnamefont
  {Boyajian}}, \bibinfo {author} {\bibfnamefont {K.}~\bibnamefont {von Braun}},
  \bibinfo {author} {\bibfnamefont {G.~A.}\ \bibnamefont {Feiden}}, \bibinfo
  {author} {\bibfnamefont {D.}~\bibnamefont {Huber}}, \bibinfo {author}
  {\bibfnamefont {S.}~\bibnamefont {Basu}}, \bibinfo {author} {\bibfnamefont
  {P.}~\bibnamefont {Demarque}}, \bibinfo {author} {\bibfnamefont {D.~A.}\
  \bibnamefont {Fischer}}, \bibinfo {author} {\bibfnamefont {G.}~\bibnamefont
  {Schaefer}}, \bibinfo {author} {\bibfnamefont {A.~W.}\ \bibnamefont {Mann}},
  \bibinfo {author} {\bibfnamefont {T.~R.}\ \bibnamefont {White}}, \emph
  {et~al.},\ }\bibfield  {title} {\bibinfo {title} {Stellar diameters and
  temperatures--vi. high angular resolution measurements of the transiting
  exoplanet host stars hd 189733 and hd 209458 and implications for models of
  cool dwarfs},\ }\href@noop {} {\bibfield  {journal} {\bibinfo  {journal}
  {Monthly Notices of the Royal Astronomical Society}\ }\textbf {\bibinfo
  {volume} {447}},\ \bibinfo {pages} {846} (\bibinfo {year}
  {2015})}\BibitemShut {NoStop}%
\bibitem [{\citenamefont {Pan}\ \emph {et~al.}(2023)\citenamefont {Pan},
  \citenamefont {Song},\ and\ \citenamefont {Long}}]{pan2023free}%
  \BibitemOpen
  \bibfield  {author} {\bibinfo {author} {\bibfnamefont {D.}~\bibnamefont
  {Pan}}, \bibinfo {author} {\bibfnamefont {X.-T.}\ \bibnamefont {Song}},\ and\
  \bibinfo {author} {\bibfnamefont {G.-L.}\ \bibnamefont {Long}},\ }\bibfield
  {title} {\bibinfo {title} {Free-space quantum secure direct communication:
  Basics, progress, and outlook},\ }\href@noop {} {\bibfield  {journal}
  {\bibinfo  {journal} {Advanced Devices \& Instrumentation}\ }\textbf
  {\bibinfo {volume} {4}},\ \bibinfo {pages} {0004} (\bibinfo {year}
  {2023})}\BibitemShut {NoStop}%
\bibitem [{\citenamefont {Bloom}\ \emph {et~al.}(2003)\citenamefont {Bloom},
  \citenamefont {Korevaar}, \citenamefont {Schuster},\ and\ \citenamefont
  {Willebrand}}]{bloom2003understanding}%
  \BibitemOpen
  \bibfield  {author} {\bibinfo {author} {\bibfnamefont {S.}~\bibnamefont
  {Bloom}}, \bibinfo {author} {\bibfnamefont {E.}~\bibnamefont {Korevaar}},
  \bibinfo {author} {\bibfnamefont {J.}~\bibnamefont {Schuster}},\ and\
  \bibinfo {author} {\bibfnamefont {H.}~\bibnamefont {Willebrand}},\ }\bibfield
   {title} {\bibinfo {title} {Understanding the performance of free-space
  optics},\ }\href@noop {} {\bibfield  {journal} {\bibinfo  {journal} {Journal
  of optical Networking}\ }\textbf {\bibinfo {volume} {2}},\ \bibinfo {pages}
  {178} (\bibinfo {year} {2003})}\BibitemShut {NoStop}%
\bibitem [{\citenamefont {Khalighi}\ and\ \citenamefont
  {Uysal}(2014)}]{khalighi2014survey}%
  \BibitemOpen
  \bibfield  {author} {\bibinfo {author} {\bibfnamefont {M.~A.}\ \bibnamefont
  {Khalighi}}\ and\ \bibinfo {author} {\bibfnamefont {M.}~\bibnamefont
  {Uysal}},\ }\bibfield  {title} {\bibinfo {title} {Survey on free space
  optical communication: A communication theory perspective},\ }\href@noop {}
  {\bibfield  {journal} {\bibinfo  {journal} {IEEE communications surveys \&
  tutorials}\ }\textbf {\bibinfo {volume} {16}},\ \bibinfo {pages} {2231}
  (\bibinfo {year} {2014})}\BibitemShut {NoStop}%
\bibitem [{\citenamefont {Luo}\ \emph {et~al.}(2019)\citenamefont {Luo},
  \citenamefont {Zhong}, \citenamefont {Erhard}, \citenamefont {Wang},
  \citenamefont {Peng}, \citenamefont {Krenn}, \citenamefont {Jiang},
  \citenamefont {Li}, \citenamefont {Liu}, \citenamefont {Lu} \emph
  {et~al.}}]{luo2019quantum}%
  \BibitemOpen
  \bibfield  {author} {\bibinfo {author} {\bibfnamefont {Y.-H.}\ \bibnamefont
  {Luo}}, \bibinfo {author} {\bibfnamefont {H.-S.}\ \bibnamefont {Zhong}},
  \bibinfo {author} {\bibfnamefont {M.}~\bibnamefont {Erhard}}, \bibinfo
  {author} {\bibfnamefont {X.-L.}\ \bibnamefont {Wang}}, \bibinfo {author}
  {\bibfnamefont {L.-C.}\ \bibnamefont {Peng}}, \bibinfo {author}
  {\bibfnamefont {M.}~\bibnamefont {Krenn}}, \bibinfo {author} {\bibfnamefont
  {X.}~\bibnamefont {Jiang}}, \bibinfo {author} {\bibfnamefont
  {L.}~\bibnamefont {Li}}, \bibinfo {author} {\bibfnamefont {N.-L.}\
  \bibnamefont {Liu}}, \bibinfo {author} {\bibfnamefont {C.-Y.}\ \bibnamefont
  {Lu}}, \emph {et~al.},\ }\bibfield  {title} {\bibinfo {title} {Quantum
  teleportation in high dimensions},\ }\href@noop {} {\bibfield  {journal}
  {\bibinfo  {journal} {Phys. Rev. Lett.}\ }\textbf {\bibinfo {volume} {123}},\
  \bibinfo {pages} {070505} (\bibinfo {year} {2019})}\BibitemShut {NoStop}%
\bibitem [{\citenamefont {Arnold}\ \emph {et~al.}(2024)\citenamefont {Arnold},
  \citenamefont {Lualdi}, \citenamefont {Goggin},\ and\ \citenamefont
  {Kwiat}}]{Nathan2024}%
  \BibitemOpen
  \bibfield  {author} {\bibinfo {author} {\bibfnamefont {N.~T.}\ \bibnamefont
  {Arnold}}, \bibinfo {author} {\bibfnamefont {C.~P.}\ \bibnamefont {Lualdi}},
  \bibinfo {author} {\bibfnamefont {M.~E.}\ \bibnamefont {Goggin}},\ and\
  \bibinfo {author} {\bibfnamefont {P.~G.}\ \bibnamefont {Kwiat}},\ }\bibfield
  {title} {\bibinfo {title} {{All-optical quantum memory}},\ }in\ \href
  {https://doi.org/10.1117/12.3003228} {\emph {\bibinfo {booktitle} {Quantum
  Computing, Communication, and Simulation IV}}},\ Vol.\ \bibinfo {volume}
  {12911},\ \bibinfo {editor} {edited by\ \bibinfo {editor} {\bibfnamefont
  {P.~R.}\ \bibnamefont {Hemmer}}\ and\ \bibinfo {editor} {\bibfnamefont
  {A.~L.}\ \bibnamefont {Migdall}}},\ \bibinfo {organization} {International
  Society for Optics and Photonics}\ (\bibinfo  {publisher} {SPIE},\ \bibinfo
  {year} {2024})\ p.\ \bibinfo {pages} {129111C}\BibitemShut {NoStop}%
\bibitem [{\citenamefont {Keiser}(2000)}]{keiser2000optical}%
  \BibitemOpen
  \bibfield  {author} {\bibinfo {author} {\bibfnamefont {G.}~\bibnamefont
  {Keiser}},\ }\href@noop {} {\emph {\bibinfo {title} {Optical fiber
  communications}}},\ Vol.~\bibinfo {volume} {2}\ (\bibinfo  {publisher}
  {McGraw-Hill New York},\ \bibinfo {year} {2000})\BibitemShut {NoStop}%
\bibitem [{\citenamefont {Hasegawa}\ \emph {et~al.}(2018)\citenamefont
  {Hasegawa}, \citenamefont {Tamura}, \citenamefont {Sakuma}, \citenamefont
  {Kawaguchi}, \citenamefont {Yamamoto},\ and\ \citenamefont
  {Koyano}}]{hasegawa2018first}%
  \BibitemOpen
  \bibfield  {author} {\bibinfo {author} {\bibfnamefont {T.}~\bibnamefont
  {Hasegawa}}, \bibinfo {author} {\bibfnamefont {Y.}~\bibnamefont {Tamura}},
  \bibinfo {author} {\bibfnamefont {H.}~\bibnamefont {Sakuma}}, \bibinfo
  {author} {\bibfnamefont {Y.}~\bibnamefont {Kawaguchi}}, \bibinfo {author}
  {\bibfnamefont {Y.}~\bibnamefont {Yamamoto}},\ and\ \bibinfo {author}
  {\bibfnamefont {Y.}~\bibnamefont {Koyano}},\ }\bibfield  {title} {\bibinfo
  {title} {The first 0.14-db/km ultra-low loss optical fiber},\ }\href@noop {}
  {\bibfield  {journal} {\bibinfo  {journal} {SEI Tech. Rev}\ }\textbf
  {\bibinfo {volume} {86}},\ \bibinfo {pages} {18} (\bibinfo {year}
  {2018})}\BibitemShut {NoStop}%
\bibitem [{\citenamefont {Ralph}\ and\ \citenamefont
  {Lund}(2009)}]{ralph2009nondeterministic}%
  \BibitemOpen
  \bibfield  {author} {\bibinfo {author} {\bibfnamefont {T.}~\bibnamefont
  {Ralph}}\ and\ \bibinfo {author} {\bibfnamefont {A.}~\bibnamefont {Lund}},\
  }\bibfield  {title} {\bibinfo {title} {Nondeterministic noiseless linear
  amplification of quantum systems},\ }in\ \href@noop {} {\emph {\bibinfo
  {booktitle} {AIP Conference Proceedings}}},\ Vol.\ \bibinfo {volume} {1110}\
  (\bibinfo {organization} {American Institute of Physics},\ \bibinfo {year}
  {2009})\ pp.\ \bibinfo {pages} {155--160}\BibitemShut {NoStop}%
\bibitem [{\citenamefont {Ralph}(2011)}]{ralph2011quantum}%
  \BibitemOpen
  \bibfield  {author} {\bibinfo {author} {\bibfnamefont {T.}~\bibnamefont
  {Ralph}},\ }\bibfield  {title} {\bibinfo {title} {Quantum error correction of
  continuous-variable states against gaussian noise},\ }\href@noop {}
  {\bibfield  {journal} {\bibinfo  {journal} {Phys. Rev. A}\ }\textbf {\bibinfo
  {volume} {84}},\ \bibinfo {pages} {022339} (\bibinfo {year}
  {2011})}\BibitemShut {NoStop}%
\bibitem [{\citenamefont {Dias}\ and\ \citenamefont
  {Ralph}(2017)}]{dias2017quantum}%
  \BibitemOpen
  \bibfield  {author} {\bibinfo {author} {\bibfnamefont {J.}~\bibnamefont
  {Dias}}\ and\ \bibinfo {author} {\bibfnamefont {T.~C.}\ \bibnamefont
  {Ralph}},\ }\bibfield  {title} {\bibinfo {title} {Quantum repeaters using
  continuous-variable teleportation},\ }\href@noop {} {\bibfield  {journal}
  {\bibinfo  {journal} {Phys. Rev. A}\ }\textbf {\bibinfo {volume} {95}},\
  \bibinfo {pages} {022312} (\bibinfo {year} {2017})}\BibitemShut {NoStop}%
\bibitem [{\citenamefont {Furrer}\ and\ \citenamefont
  {Munro}(2018)}]{furrer2018repeaters}%
  \BibitemOpen
  \bibfield  {author} {\bibinfo {author} {\bibfnamefont {F.}~\bibnamefont
  {Furrer}}\ and\ \bibinfo {author} {\bibfnamefont {W.~J.}\ \bibnamefont
  {Munro}},\ }\bibfield  {title} {\bibinfo {title} {Repeaters for
  continuous-variable quantum communication},\ }\href@noop {} {\bibfield
  {journal} {\bibinfo  {journal} {Phys. Rev. A}\ }\textbf {\bibinfo {volume}
  {98}},\ \bibinfo {pages} {032335} (\bibinfo {year} {2018})}\BibitemShut
  {NoStop}%
\bibitem [{\citenamefont {Liu}\ \emph {et~al.}(2018)\citenamefont {Liu},
  \citenamefont {Yan}, \citenamefont {Ma}, \citenamefont {Yan},\ and\
  \citenamefont {Jia}}]{liu2018continuous}%
  \BibitemOpen
  \bibfield  {author} {\bibinfo {author} {\bibfnamefont {Y.}~\bibnamefont
  {Liu}}, \bibinfo {author} {\bibfnamefont {J.}~\bibnamefont {Yan}}, \bibinfo
  {author} {\bibfnamefont {L.}~\bibnamefont {Ma}}, \bibinfo {author}
  {\bibfnamefont {Z.}~\bibnamefont {Yan}},\ and\ \bibinfo {author}
  {\bibfnamefont {X.}~\bibnamefont {Jia}},\ }\bibfield  {title} {\bibinfo
  {title} {Continuous-variable entanglement distillation between remote quantum
  nodes},\ }\href@noop {} {\bibfield  {journal} {\bibinfo  {journal} {Phys.
  Rev. A}\ }\textbf {\bibinfo {volume} {98}},\ \bibinfo {pages} {052308}
  (\bibinfo {year} {2018})}\BibitemShut {NoStop}%
\bibitem [{\citenamefont {Dias}\ \emph {et~al.}(2020)\citenamefont {Dias},
  \citenamefont {Winnel}, \citenamefont {Hosseinidehaj},\ and\ \citenamefont
  {Ralph}}]{dias2020quantum}%
  \BibitemOpen
  \bibfield  {author} {\bibinfo {author} {\bibfnamefont {J.}~\bibnamefont
  {Dias}}, \bibinfo {author} {\bibfnamefont {M.~S.}\ \bibnamefont {Winnel}},
  \bibinfo {author} {\bibfnamefont {N.}~\bibnamefont {Hosseinidehaj}},\ and\
  \bibinfo {author} {\bibfnamefont {T.~C.}\ \bibnamefont {Ralph}},\ }\bibfield
  {title} {\bibinfo {title} {Quantum repeater for continuous-variable
  entanglement distribution},\ }\href@noop {} {\bibfield  {journal} {\bibinfo
  {journal} {Phys. Rev. A}\ }\textbf {\bibinfo {volume} {102}},\ \bibinfo
  {pages} {052425} (\bibinfo {year} {2020})}\BibitemShut {NoStop}%
\bibitem [{\citenamefont {Seshadreesan}\ \emph {et~al.}(2020)\citenamefont
  {Seshadreesan}, \citenamefont {Krovi},\ and\ \citenamefont
  {Guha}}]{seshadreesan2020continuous}%
  \BibitemOpen
  \bibfield  {author} {\bibinfo {author} {\bibfnamefont {K.~P.}\ \bibnamefont
  {Seshadreesan}}, \bibinfo {author} {\bibfnamefont {H.}~\bibnamefont
  {Krovi}},\ and\ \bibinfo {author} {\bibfnamefont {S.}~\bibnamefont {Guha}},\
  }\bibfield  {title} {\bibinfo {title} {Continuous-variable quantum repeater
  based on quantum scissors and mode multiplexing},\ }\href@noop {} {\bibfield
  {journal} {\bibinfo  {journal} {Phys. Rev. Res.}\ }\textbf {\bibinfo {volume}
  {2}},\ \bibinfo {pages} {013310} (\bibinfo {year} {2020})}\BibitemShut
  {NoStop}%
\bibitem [{\citenamefont {Azuma}\ \emph {et~al.}(2023)\citenamefont {Azuma},
  \citenamefont {Economou}, \citenamefont {Elkouss}, \citenamefont {Hilaire},
  \citenamefont {Jiang}, \citenamefont {Lo},\ and\ \citenamefont
  {Tzitrin}}]{Azuma2023}%
  \BibitemOpen
  \bibfield  {author} {\bibinfo {author} {\bibfnamefont {K.}~\bibnamefont
  {Azuma}}, \bibinfo {author} {\bibfnamefont {S.~E.}\ \bibnamefont {Economou}},
  \bibinfo {author} {\bibfnamefont {D.}~\bibnamefont {Elkouss}}, \bibinfo
  {author} {\bibfnamefont {P.}~\bibnamefont {Hilaire}}, \bibinfo {author}
  {\bibfnamefont {L.}~\bibnamefont {Jiang}}, \bibinfo {author} {\bibfnamefont
  {H.-K.}\ \bibnamefont {Lo}},\ and\ \bibinfo {author} {\bibfnamefont
  {I.}~\bibnamefont {Tzitrin}},\ }\bibfield  {title} {\bibinfo {title} {Quantum
  repeaters: From quantum networks to the quantum internet},\ }\href
  {https://doi.org/10.1103/RevModPhys.95.045006} {\bibfield  {journal}
  {\bibinfo  {journal} {Rev. Mod. Phys.}\ }\textbf {\bibinfo {volume} {95}},\
  \bibinfo {pages} {045006} (\bibinfo {year} {2023})}\BibitemShut {NoStop}%
\bibitem [{\citenamefont {Foreman}\ \emph {et~al.}(2007)\citenamefont
  {Foreman}, \citenamefont {Holman}, \citenamefont {Hudson}, \citenamefont
  {Jones},\ and\ \citenamefont {Ye}}]{foreman2007remote}%
  \BibitemOpen
  \bibfield  {author} {\bibinfo {author} {\bibfnamefont {S.~M.}\ \bibnamefont
  {Foreman}}, \bibinfo {author} {\bibfnamefont {K.~W.}\ \bibnamefont {Holman}},
  \bibinfo {author} {\bibfnamefont {D.~D.}\ \bibnamefont {Hudson}}, \bibinfo
  {author} {\bibfnamefont {D.~J.}\ \bibnamefont {Jones}},\ and\ \bibinfo
  {author} {\bibfnamefont {J.}~\bibnamefont {Ye}},\ }\bibfield  {title}
  {\bibinfo {title} {Remote transfer of ultrastable frequency references via
  fiber networks},\ }\href@noop {} {\bibfield  {journal} {\bibinfo  {journal}
  {Review of Scientific Instruments}\ }\textbf {\bibinfo {volume} {78}},\
  \bibinfo {pages} {021101} (\bibinfo {year} {2007})}\BibitemShut {NoStop}%
\bibitem [{\citenamefont {{\"O}zdemir}\ \emph {et~al.}(2001)\citenamefont
  {{\"O}zdemir}, \citenamefont {Miranowicz}, \citenamefont {Koashi},\ and\
  \citenamefont {Imoto}}]{ozdemir2001quantum}%
  \BibitemOpen
  \bibfield  {author} {\bibinfo {author} {\bibfnamefont {{\c{S}}.~K.}\
  \bibnamefont {{\"O}zdemir}}, \bibinfo {author} {\bibfnamefont
  {A.}~\bibnamefont {Miranowicz}}, \bibinfo {author} {\bibfnamefont
  {M.}~\bibnamefont {Koashi}},\ and\ \bibinfo {author} {\bibfnamefont
  {N.}~\bibnamefont {Imoto}},\ }\bibfield  {title} {\bibinfo {title}
  {Quantum-scissors device for optical state truncation: A proposal for
  practical realization},\ }\href@noop {} {\bibfield  {journal} {\bibinfo
  {journal} {Phys. Rev. A}\ }\textbf {\bibinfo {volume} {64}},\ \bibinfo
  {pages} {063818} (\bibinfo {year} {2001})}\BibitemShut {NoStop}%
\bibitem [{\citenamefont {Pegg}\ \emph {et~al.}(1998)\citenamefont {Pegg},
  \citenamefont {Phillips},\ and\ \citenamefont {Barnett}}]{pegg1998optical}%
  \BibitemOpen
  \bibfield  {author} {\bibinfo {author} {\bibfnamefont {D.~T.}\ \bibnamefont
  {Pegg}}, \bibinfo {author} {\bibfnamefont {L.~S.}\ \bibnamefont {Phillips}},\
  and\ \bibinfo {author} {\bibfnamefont {S.~M.}\ \bibnamefont {Barnett}},\
  }\bibfield  {title} {\bibinfo {title} {Optical state truncation by projection
  synthesis},\ }\href@noop {} {\bibfield  {journal} {\bibinfo  {journal} {Phys.
  Rev. Lett.}\ }\textbf {\bibinfo {volume} {81}},\ \bibinfo {pages} {1604}
  (\bibinfo {year} {1998})}\BibitemShut {NoStop}%
\bibitem [{\citenamefont {Lund}\ and\ \citenamefont
  {Ralph}(2009)}]{lund2009continuous}%
  \BibitemOpen
  \bibfield  {author} {\bibinfo {author} {\bibfnamefont {A.}~\bibnamefont
  {Lund}}\ and\ \bibinfo {author} {\bibfnamefont {T.}~\bibnamefont {Ralph}},\
  }\bibfield  {title} {\bibinfo {title} {Continuous-variable entanglement
  distillation over a general lossy channel},\ }\href@noop {} {\bibfield
  {journal} {\bibinfo  {journal} {Phys. Rev. A}\ }\textbf {\bibinfo {volume}
  {80}},\ \bibinfo {pages} {032309} (\bibinfo {year} {2009})}\BibitemShut
  {NoStop}%
\bibitem [{\citenamefont
  {Fiur{\'a}{\v{s}}ek}(2010)}]{fiuravsek2010distillation}%
  \BibitemOpen
  \bibfield  {author} {\bibinfo {author} {\bibfnamefont {J.}~\bibnamefont
  {Fiur{\'a}{\v{s}}ek}},\ }\bibfield  {title} {\bibinfo {title} {Distillation
  and purification of symmetric entangled gaussian states},\ }\href@noop {}
  {\bibfield  {journal} {\bibinfo  {journal} {Phys. Rev. A}\ }\textbf {\bibinfo
  {volume} {82}},\ \bibinfo {pages} {042331} (\bibinfo {year}
  {2010})}\BibitemShut {NoStop}%
\bibitem [{\citenamefont {Browne}\ \emph {et~al.}(2003)\citenamefont {Browne},
  \citenamefont {Eisert}, \citenamefont {Scheel},\ and\ \citenamefont
  {Plenio}}]{browne2003driving}%
  \BibitemOpen
  \bibfield  {author} {\bibinfo {author} {\bibfnamefont {D.~E.}\ \bibnamefont
  {Browne}}, \bibinfo {author} {\bibfnamefont {J.}~\bibnamefont {Eisert}},
  \bibinfo {author} {\bibfnamefont {S.}~\bibnamefont {Scheel}},\ and\ \bibinfo
  {author} {\bibfnamefont {M.~B.}\ \bibnamefont {Plenio}},\ }\bibfield  {title}
  {\bibinfo {title} {Driving non-gaussian to gaussian states with linear
  optics},\ }\href@noop {} {\bibfield  {journal} {\bibinfo  {journal} {Phys.
  Rev. A}\ }\textbf {\bibinfo {volume} {67}},\ \bibinfo {pages} {062320}
  (\bibinfo {year} {2003})}\BibitemShut {NoStop}%
\bibitem [{\citenamefont {Eisert}\ \emph {et~al.}(2004)\citenamefont {Eisert},
  \citenamefont {Browne}, \citenamefont {Scheel},\ and\ \citenamefont
  {Plenio}}]{eisert2004distillation}%
  \BibitemOpen
  \bibfield  {author} {\bibinfo {author} {\bibfnamefont {J.}~\bibnamefont
  {Eisert}}, \bibinfo {author} {\bibfnamefont {D.}~\bibnamefont {Browne}},
  \bibinfo {author} {\bibfnamefont {S.}~\bibnamefont {Scheel}},\ and\ \bibinfo
  {author} {\bibfnamefont {M.}~\bibnamefont {Plenio}},\ }\bibfield  {title}
  {\bibinfo {title} {Distillation of continuous-variable entanglement with
  optical means},\ }\href@noop {} {\bibfield  {journal} {\bibinfo  {journal}
  {Annals of Physics}\ }\textbf {\bibinfo {volume} {311}},\ \bibinfo {pages}
  {431} (\bibinfo {year} {2004})}\BibitemShut {NoStop}%
\bibitem [{\citenamefont {Rozp{\k{e}}dek}\ \emph {et~al.}(2021)\citenamefont
  {Rozp{\k{e}}dek}, \citenamefont {Noh}, \citenamefont {Xu}, \citenamefont
  {Guha},\ and\ \citenamefont {Jiang}}]{rozpkedek2021quantum}%
  \BibitemOpen
  \bibfield  {author} {\bibinfo {author} {\bibfnamefont {F.}~\bibnamefont
  {Rozp{\k{e}}dek}}, \bibinfo {author} {\bibfnamefont {K.}~\bibnamefont {Noh}},
  \bibinfo {author} {\bibfnamefont {Q.}~\bibnamefont {Xu}}, \bibinfo {author}
  {\bibfnamefont {S.}~\bibnamefont {Guha}},\ and\ \bibinfo {author}
  {\bibfnamefont {L.}~\bibnamefont {Jiang}},\ }\bibfield  {title} {\bibinfo
  {title} {Quantum repeaters based on concatenated bosonic and
  discrete-variable quantum codes},\ }\href@noop {} {\bibfield  {journal}
  {\bibinfo  {journal} {npj Quantum Information}\ }\textbf {\bibinfo {volume}
  {7}},\ \bibinfo {pages} {102} (\bibinfo {year} {2021})}\BibitemShut {NoStop}%
\bibitem [{\citenamefont {Lassen}\ \emph
  {et~al.}(2010{\natexlab{a}})\citenamefont {Lassen}, \citenamefont {Sabuncu},
  \citenamefont {Huck}, \citenamefont {Niset}, \citenamefont {Leuchs},
  \citenamefont {Cerf},\ and\ \citenamefont {Andersen}}]{lassen2010quantum}%
  \BibitemOpen
  \bibfield  {author} {\bibinfo {author} {\bibfnamefont {M.}~\bibnamefont
  {Lassen}}, \bibinfo {author} {\bibfnamefont {M.}~\bibnamefont {Sabuncu}},
  \bibinfo {author} {\bibfnamefont {A.}~\bibnamefont {Huck}}, \bibinfo {author}
  {\bibfnamefont {J.}~\bibnamefont {Niset}}, \bibinfo {author} {\bibfnamefont
  {G.}~\bibnamefont {Leuchs}}, \bibinfo {author} {\bibfnamefont {N.~J.}\
  \bibnamefont {Cerf}},\ and\ \bibinfo {author} {\bibfnamefont {U.~L.}\
  \bibnamefont {Andersen}},\ }\bibfield  {title} {\bibinfo {title} {Quantum
  optical coherence can survive photon losses using a continuous-variable
  quantum erasure-correcting code},\ }\href@noop {} {\bibfield  {journal}
  {\bibinfo  {journal} {Nature Photonics}\ }\textbf {\bibinfo {volume} {4}},\
  \bibinfo {pages} {700} (\bibinfo {year} {2010}{\natexlab{a}})}\BibitemShut
  {NoStop}%
\bibitem [{\citenamefont {Lassen}\ \emph {et~al.}(2013)\citenamefont {Lassen},
  \citenamefont {Berni}, \citenamefont {Madsen}, \citenamefont {Filip},\ and\
  \citenamefont {Andersen}}]{lassen2013gaussian}%
  \BibitemOpen
  \bibfield  {author} {\bibinfo {author} {\bibfnamefont {M.}~\bibnamefont
  {Lassen}}, \bibinfo {author} {\bibfnamefont {A.}~\bibnamefont {Berni}},
  \bibinfo {author} {\bibfnamefont {L.~S.}\ \bibnamefont {Madsen}}, \bibinfo
  {author} {\bibfnamefont {R.}~\bibnamefont {Filip}},\ and\ \bibinfo {author}
  {\bibfnamefont {U.~L.}\ \bibnamefont {Andersen}},\ }\bibfield  {title}
  {\bibinfo {title} {Gaussian error correction of quantum states in a
  correlated noisy channel},\ }\href@noop {} {\bibfield  {journal} {\bibinfo
  {journal} {Phys. Rev. Lett.}\ }\textbf {\bibinfo {volume} {111}},\ \bibinfo
  {pages} {180502} (\bibinfo {year} {2013})}\BibitemShut {NoStop}%
\bibitem [{\citenamefont {Lassen}\ \emph {et~al.}(2007)\citenamefont {Lassen},
  \citenamefont {Sabuncu}, \citenamefont {Buchhave},\ and\ \citenamefont
  {Andersen}}]{lassen2007generation}%
  \BibitemOpen
  \bibfield  {author} {\bibinfo {author} {\bibfnamefont {M.}~\bibnamefont
  {Lassen}}, \bibinfo {author} {\bibfnamefont {M.}~\bibnamefont {Sabuncu}},
  \bibinfo {author} {\bibfnamefont {P.}~\bibnamefont {Buchhave}},\ and\
  \bibinfo {author} {\bibfnamefont {U.~L.}\ \bibnamefont {Andersen}},\
  }\bibfield  {title} {\bibinfo {title} {Generation of polarization squeezing
  with periodically poled ktp at 1064 nm},\ }\href@noop {} {\bibfield
  {journal} {\bibinfo  {journal} {Optics Express}\ }\textbf {\bibinfo {volume}
  {15}},\ \bibinfo {pages} {5077} (\bibinfo {year} {2007})}\BibitemShut
  {NoStop}%
\bibitem [{\citenamefont {Lassen}\ \emph
  {et~al.}(2010{\natexlab{b}})\citenamefont {Lassen}, \citenamefont {Madsen},
  \citenamefont {Sabuncu}, \citenamefont {Filip},\ and\ \citenamefont
  {Andersen}}]{lassen2010experimental}%
  \BibitemOpen
  \bibfield  {author} {\bibinfo {author} {\bibfnamefont {M.}~\bibnamefont
  {Lassen}}, \bibinfo {author} {\bibfnamefont {L.~S.}\ \bibnamefont {Madsen}},
  \bibinfo {author} {\bibfnamefont {M.}~\bibnamefont {Sabuncu}}, \bibinfo
  {author} {\bibfnamefont {R.}~\bibnamefont {Filip}},\ and\ \bibinfo {author}
  {\bibfnamefont {U.~L.}\ \bibnamefont {Andersen}},\ }\bibfield  {title}
  {\bibinfo {title} {Experimental demonstration of squeezed-state quantum
  averaging},\ }\href@noop {} {\bibfield  {journal} {\bibinfo  {journal} {Phys.
  Rev. A—Atomic, Molecular, and Optical Physics}\ }\textbf {\bibinfo {volume}
  {82}},\ \bibinfo {pages} {021801} (\bibinfo {year}
  {2010}{\natexlab{b}})}\BibitemShut {NoStop}%
\bibitem [{\citenamefont {Vahlbruch}\ \emph {et~al.}(2016)\citenamefont
  {Vahlbruch}, \citenamefont {Mehmet}, \citenamefont {Danzmann},\ and\
  \citenamefont {Schnabel}}]{vahlbruch2016detection}%
  \BibitemOpen
  \bibfield  {author} {\bibinfo {author} {\bibfnamefont {H.}~\bibnamefont
  {Vahlbruch}}, \bibinfo {author} {\bibfnamefont {M.}~\bibnamefont {Mehmet}},
  \bibinfo {author} {\bibfnamefont {K.}~\bibnamefont {Danzmann}},\ and\
  \bibinfo {author} {\bibfnamefont {R.}~\bibnamefont {Schnabel}},\ }\bibfield
  {title} {\bibinfo {title} {Detection of 15 db squeezed states of light and
  their application for the absolute calibration of photoelectric quantum
  efficiency},\ }\href@noop {} {\bibfield  {journal} {\bibinfo  {journal}
  {Phys. Rev. Lett.}\ }\textbf {\bibinfo {volume} {117}},\ \bibinfo {pages}
  {110801} (\bibinfo {year} {2016})}\BibitemShut {NoStop}%
\bibitem [{\citenamefont {Andersen}\ \emph {et~al.}(2016)\citenamefont
  {Andersen}, \citenamefont {Gehring}, \citenamefont {Marquardt},\ and\
  \citenamefont {Leuchs}}]{andersen201630}%
  \BibitemOpen
  \bibfield  {author} {\bibinfo {author} {\bibfnamefont {U.~L.}\ \bibnamefont
  {Andersen}}, \bibinfo {author} {\bibfnamefont {T.}~\bibnamefont {Gehring}},
  \bibinfo {author} {\bibfnamefont {C.}~\bibnamefont {Marquardt}},\ and\
  \bibinfo {author} {\bibfnamefont {G.}~\bibnamefont {Leuchs}},\ }\bibfield
  {title} {\bibinfo {title} {30 years of squeezed light generation},\
  }\href@noop {} {\bibfield  {journal} {\bibinfo  {journal} {Phys. Scr.}\
  }\textbf {\bibinfo {volume} {91}},\ \bibinfo {pages} {053001} (\bibinfo
  {year} {2016})}\BibitemShut {NoStop}%
\bibitem [{\citenamefont {Amari}\ \emph {et~al.}(2023)\citenamefont {Amari},
  \citenamefont {Takai},\ and\ \citenamefont {Hirano}}]{amari2023highly}%
  \BibitemOpen
  \bibfield  {author} {\bibinfo {author} {\bibfnamefont {J.}~\bibnamefont
  {Amari}}, \bibinfo {author} {\bibfnamefont {J.}~\bibnamefont {Takai}},\ and\
  \bibinfo {author} {\bibfnamefont {T.}~\bibnamefont {Hirano}},\ }\bibfield
  {title} {\bibinfo {title} {Highly efficient measurement of optical quadrature
  squeezing using a spatial light modulator controlled by machine learning},\
  }\href@noop {} {\bibfield  {journal} {\bibinfo  {journal} {Optics Continuum}\
  }\textbf {\bibinfo {volume} {2}},\ \bibinfo {pages} {933} (\bibinfo {year}
  {2023})}\BibitemShut {NoStop}%
\bibitem [{\citenamefont {Dias}\ \emph {et~al.}(2022)\citenamefont {Dias},
  \citenamefont {Winnel}, \citenamefont {Munro}, \citenamefont {Ralph},\ and\
  \citenamefont {Nemoto}}]{dias2022distributing}%
  \BibitemOpen
  \bibfield  {author} {\bibinfo {author} {\bibfnamefont {J.}~\bibnamefont
  {Dias}}, \bibinfo {author} {\bibfnamefont {M.~S.}\ \bibnamefont {Winnel}},
  \bibinfo {author} {\bibfnamefont {W.~J.}\ \bibnamefont {Munro}}, \bibinfo
  {author} {\bibfnamefont {T.}~\bibnamefont {Ralph}},\ and\ \bibinfo {author}
  {\bibfnamefont {K.}~\bibnamefont {Nemoto}},\ }\bibfield  {title} {\bibinfo
  {title} {Distributing entanglement in first-generation discrete-and
  continuous-variable quantum repeaters},\ }\href@noop {} {\bibfield  {journal}
  {\bibinfo  {journal} {Phys. Rev. A}\ }\textbf {\bibinfo {volume} {106}},\
  \bibinfo {pages} {052604} (\bibinfo {year} {2022})}\BibitemShut {NoStop}%
\bibitem [{\citenamefont {Brown}\ and\ \citenamefont
  {Twiss}(1956)}]{brown1956correlation}%
  \BibitemOpen
  \bibfield  {author} {\bibinfo {author} {\bibfnamefont {R.~H.}\ \bibnamefont
  {Brown}}\ and\ \bibinfo {author} {\bibfnamefont {R.~Q.}\ \bibnamefont
  {Twiss}},\ }\bibfield  {title} {\bibinfo {title} {Correlation between photons
  in two coherent beams of light},\ }\href@noop {} {\bibfield  {journal}
  {\bibinfo  {journal} {Nature}\ }\textbf {\bibinfo {volume} {177}},\ \bibinfo
  {pages} {27} (\bibinfo {year} {1956})}\BibitemShut {NoStop}%
\bibitem [{\citenamefont {Hale}\ \emph {et~al.}(2000)\citenamefont {Hale},
  \citenamefont {Bester}, \citenamefont {Danchi}, \citenamefont {Fitelson},
  \citenamefont {Hoss}, \citenamefont {Lipman}, \citenamefont {Monnier},
  \citenamefont {Tuthill},\ and\ \citenamefont {Townes}}]{hale2000berkeley}%
  \BibitemOpen
  \bibfield  {author} {\bibinfo {author} {\bibfnamefont {D.~D.}\ \bibnamefont
  {Hale}}, \bibinfo {author} {\bibfnamefont {M.}~\bibnamefont {Bester}},
  \bibinfo {author} {\bibfnamefont {W.}~\bibnamefont {Danchi}}, \bibinfo
  {author} {\bibfnamefont {W.}~\bibnamefont {Fitelson}}, \bibinfo {author}
  {\bibfnamefont {S.}~\bibnamefont {Hoss}}, \bibinfo {author} {\bibfnamefont
  {E.}~\bibnamefont {Lipman}}, \bibinfo {author} {\bibfnamefont
  {J.}~\bibnamefont {Monnier}}, \bibinfo {author} {\bibfnamefont
  {P.}~\bibnamefont {Tuthill}},\ and\ \bibinfo {author} {\bibfnamefont
  {C.}~\bibnamefont {Townes}},\ }\bibfield  {title} {\bibinfo {title} {The
  berkeley infrared spatial interferometer: a heterodyne stellar interferometer
  for the mid-infrared},\ }\href@noop {} {\bibfield  {journal} {\bibinfo
  {journal} {The Astrophysical Journal}\ }\textbf {\bibinfo {volume} {537}},\
  \bibinfo {pages} {998} (\bibinfo {year} {2000})}\BibitemShut {NoStop}%
\bibitem [{\citenamefont {Caves}(1981)}]{caves1981quantum}%
  \BibitemOpen
  \bibfield  {author} {\bibinfo {author} {\bibfnamefont {C.~M.}\ \bibnamefont
  {Caves}},\ }\bibfield  {title} {\bibinfo {title} {Quantum-mechanical noise in
  an interferometer},\ }\href@noop {} {\bibfield  {journal} {\bibinfo
  {journal} {Physical Review D}\ }\textbf {\bibinfo {volume} {23}},\ \bibinfo
  {pages} {1693} (\bibinfo {year} {1981})}\BibitemShut {NoStop}%
\bibitem [{\citenamefont {Purvis}\ \emph {et~al.}(2024)\citenamefont {Purvis},
  \citenamefont {Lafler},\ and\ \citenamefont {Lanning}}]{purvis2024practical}%
  \BibitemOpen
  \bibfield  {author} {\bibinfo {author} {\bibfnamefont {B.}~\bibnamefont
  {Purvis}}, \bibinfo {author} {\bibfnamefont {R.}~\bibnamefont {Lafler}},\
  and\ \bibinfo {author} {\bibfnamefont {R.~N.}\ \bibnamefont {Lanning}},\
  }\bibfield  {title} {\bibinfo {title} {Practical approach to extending
  baselines of telescopes using continuous-variable quantum information},\
  }\href@noop {} {\bibfield  {journal} {\bibinfo  {journal} {arXiv preprint
  arXiv:2403.03491}\ } (\bibinfo {year} {2024})}\BibitemShut {NoStop}%
\bibitem [{\citenamefont {Hoelscher-Obermaier}\ and\ \citenamefont {van
  Loock}(2011)}]{hoelscher2011optimal}%
  \BibitemOpen
  \bibfield  {author} {\bibinfo {author} {\bibfnamefont {J.}~\bibnamefont
  {Hoelscher-Obermaier}}\ and\ \bibinfo {author} {\bibfnamefont
  {P.}~\bibnamefont {van Loock}},\ }\bibfield  {title} {\bibinfo {title}
  {Optimal gaussian entanglement swapping},\ }\href@noop {} {\bibfield
  {journal} {\bibinfo  {journal} {Phys. Rev. A}\ }\textbf {\bibinfo {volume}
  {83}},\ \bibinfo {pages} {012319} (\bibinfo {year} {2011})}\BibitemShut
  {NoStop}%
\bibitem [{\citenamefont {Fukui}\ \emph {et~al.}(2021)\citenamefont {Fukui},
  \citenamefont {Alexander},\ and\ \citenamefont {van Loock}}]{fukui2021all}%
  \BibitemOpen
  \bibfield  {author} {\bibinfo {author} {\bibfnamefont {K.}~\bibnamefont
  {Fukui}}, \bibinfo {author} {\bibfnamefont {R.~N.}\ \bibnamefont
  {Alexander}},\ and\ \bibinfo {author} {\bibfnamefont {P.}~\bibnamefont {van
  Loock}},\ }\bibfield  {title} {\bibinfo {title} {All-optical long-distance
  quantum communication with gottesman-kitaev-preskill qubits},\ }\href@noop {}
  {\bibfield  {journal} {\bibinfo  {journal} {Phys. Rev. Res.}\ }\textbf
  {\bibinfo {volume} {3}},\ \bibinfo {pages} {033118} (\bibinfo {year}
  {2021})}\BibitemShut {NoStop}%
\bibitem [{\citenamefont {Pearce}\ \emph {et~al.}(2017)\citenamefont {Pearce},
  \citenamefont {Campbell},\ and\ \citenamefont {Kok}}]{pearce2017optimal}%
  \BibitemOpen
  \bibfield  {author} {\bibinfo {author} {\bibfnamefont {M.~E.}\ \bibnamefont
  {Pearce}}, \bibinfo {author} {\bibfnamefont {E.~T.}\ \bibnamefont
  {Campbell}},\ and\ \bibinfo {author} {\bibfnamefont {P.}~\bibnamefont
  {Kok}},\ }\bibfield  {title} {\bibinfo {title} {Optimal quantum metrology of
  distant black bodies},\ }\href@noop {} {\bibfield  {journal} {\bibinfo
  {journal} {Quantum}\ }\textbf {\bibinfo {volume} {1}},\ \bibinfo {pages} {21}
  (\bibinfo {year} {2017})}\BibitemShut {NoStop}%
\bibitem [{\citenamefont {Yuen}\ and\ \citenamefont
  {Chan}(1983)}]{yuen1983noise}%
  \BibitemOpen
  \bibfield  {author} {\bibinfo {author} {\bibfnamefont {H.~P.}\ \bibnamefont
  {Yuen}}\ and\ \bibinfo {author} {\bibfnamefont {V.~W.}\ \bibnamefont
  {Chan}},\ }\bibfield  {title} {\bibinfo {title} {Noise in homodyne and
  heterodyne detection},\ }\href@noop {} {\bibfield  {journal} {\bibinfo
  {journal} {Optics letters}\ }\textbf {\bibinfo {volume} {8}},\ \bibinfo
  {pages} {177} (\bibinfo {year} {1983})}\BibitemShut {NoStop}%
\bibitem [{\citenamefont {Li}\ \emph {et~al.}(1999)\citenamefont {Li},
  \citenamefont {Guzun},\ and\ \citenamefont {Xiao}}]{li1999sub}%
  \BibitemOpen
  \bibfield  {author} {\bibinfo {author} {\bibfnamefont {Y.-q.}\ \bibnamefont
  {Li}}, \bibinfo {author} {\bibfnamefont {D.}~\bibnamefont {Guzun}},\ and\
  \bibinfo {author} {\bibfnamefont {M.}~\bibnamefont {Xiao}},\ }\bibfield
  {title} {\bibinfo {title} {Sub-shot-noise-limited optical heterodyne
  detection using an amplitude-squeezed local oscillator},\ }\href@noop {}
  {\bibfield  {journal} {\bibinfo  {journal} {Phys. Rev. Lett.}\ }\textbf
  {\bibinfo {volume} {82}},\ \bibinfo {pages} {5225} (\bibinfo {year}
  {1999})}\BibitemShut {NoStop}%
\bibitem [{\citenamefont {Li}\ \emph {et~al.}(2000)\citenamefont {Li},
  \citenamefont {Guzun},\ and\ \citenamefont {Xiao}}]{li2000li}%
  \BibitemOpen
  \bibfield  {author} {\bibinfo {author} {\bibfnamefont {Y.}~\bibnamefont
  {Li}}, \bibinfo {author} {\bibfnamefont {D.}~\bibnamefont {Guzun}},\ and\
  \bibinfo {author} {\bibfnamefont {M.}~\bibnamefont {Xiao}},\ }\bibfield
  {title} {\bibinfo {title} {Li, guzun, and xiao reply},\ }\href@noop {}
  {\bibfield  {journal} {\bibinfo  {journal} {Phys. Rev. Lett.}\ }\textbf
  {\bibinfo {volume} {85}},\ \bibinfo {pages} {678} (\bibinfo {year}
  {2000})}\BibitemShut {NoStop}%
\bibitem [{\citenamefont {Ralph}(2000)}]{ralph2000can}%
  \BibitemOpen
  \bibfield  {author} {\bibinfo {author} {\bibfnamefont {T.}~\bibnamefont
  {Ralph}},\ }\bibfield  {title} {\bibinfo {title} {Can signal-to-noise be
  improved by heterodyne detection using an amplitude squeezed local
  oscillator?},\ }\href@noop {} {\bibfield  {journal} {\bibinfo  {journal}
  {Phys. Rev. Lett.}\ }\textbf {\bibinfo {volume} {85}},\ \bibinfo {pages}
  {677} (\bibinfo {year} {2000})}\BibitemShut {NoStop}%
\bibitem [{\citenamefont {Tsang}\ \emph {et~al.}(2011)\citenamefont {Tsang},
  \citenamefont {Wiseman},\ and\ \citenamefont {Caves}}]{tsang2011fundamental}%
  \BibitemOpen
  \bibfield  {author} {\bibinfo {author} {\bibfnamefont {M.}~\bibnamefont
  {Tsang}}, \bibinfo {author} {\bibfnamefont {H.~M.}\ \bibnamefont {Wiseman}},\
  and\ \bibinfo {author} {\bibfnamefont {C.~M.}\ \bibnamefont {Caves}},\
  }\bibfield  {title} {\bibinfo {title} {Fundamental quantum limit to waveform
  estimation},\ }\href@noop {} {\bibfield  {journal} {\bibinfo  {journal}
  {Phys. Rev. Lett.}\ }\textbf {\bibinfo {volume} {106}},\ \bibinfo {pages}
  {090401} (\bibinfo {year} {2011})}\BibitemShut {NoStop}%
\end{thebibliography}%


\begin{thebibliography}{10}

\bibitem{zernike1938concept}
Frederik Zernike.
\newblock The concept of degree of coherence and its application to optical
  problems.
\newblock {\em Physica}, 5(8):785--795, 1938.

\bibitem{collaboration2019first}
Event Horizon~Telescope Collaboration, 
\newblock First m87 event horizon telescope results. i. the shadow of the
  supermassive black hole.
\newblock {\em Astrophys. J. Lett}, 875(1):L1, 2019.

\bibitem{wilson2009tools}
Thomas~L Wilson, Kristen Rohlfs, and Susanne H{\"u}ttemeister.
\newblock {\em Tools of radio astronomy}, volume~5.
\newblock Springer, 2009.

\bibitem{monnier2003optical}
John~D Monnier.
\newblock Optical interferometry in astronomy.
\newblock {\em Reports on Progress in Physics}, 66(5):789, 2003.

\bibitem{tsang2011quantum}
Mankei Tsang.
\newblock Quantum nonlocality in weak-thermal-light interferometry.
\newblock {\em Physical review letters}, 107(27):270402, 2011.

\bibitem{gottesman2012longer}
Daniel Gottesman, Thomas Jennewein, and Sarah Croke.
\newblock Longer-baseline telescopes using quantum repeaters.
\newblock {\em Physical review letters}, 109(7):070503, 2012.

\bibitem{khabiboulline2019optical}
Emil~T Khabiboulline, Johannes Borregaard, Kristiaan De~Greve, and Mikhail~D
  Lukin.
\newblock Optical interferometry with quantum networks.
\newblock {\em Physical review letters}, 123(7):070504, 2019.

\bibitem{khabiboulline2019quantum}
Emil~T Khabiboulline, Johannes Borregaard, Kristiaan De~Greve, and Mikhail~D
  Lukin.
\newblock Quantum-assisted telescope arrays.
\newblock {\em Physical Review A}, 100(2):022316, 2019.


\bibitem{Huang2022}
Z. Huang, G. K. Brennen, and Y. Ouyang,
``Imaging Stars with Quantum Error Correction,''
\textit{Phys. Rev. Lett.} \textbf{129}, no. 21, 210502 (2022),
\doi{10.1103/PhysRevLett.129.210502},
\url{https://link.aps.org/doi/10.1103/PhysRevLett.129.210502}.

\bibitem{Czupryniak2023}
R. Czupryniak, J. Steinmetz, P. G. Kwiat, and A. N. Jordan,
``Optimal qubit circuits for quantum-enhanced telescopes,''
\textit{Phys. Rev. A} \textbf{108}, no. 5, 052408 (2023),
\doi{10.1103/PhysRevA.108.052408},
\url{https://link.aps.org/doi/10.1103/PhysRevA.108.052408}.

\bibitem{sangouard2011quantum}
Nicolas Sangouard, Christoph Simon, Hugues De~Riedmatten, and Nicolas Gisin.
\newblock Quantum repeaters based on atomic ensembles and linear optics.
\newblock {\em Reviews of Modern Physics}, 83(1):33, 2011.

\bibitem{Czupryniak2022}
R. Czupryniak, E. Chitambar, J. Steinmetz, and A. N. Jordan,
``Quantum telescopy clock games,''
\textit{Phys. Rev. A} \textbf{106}, no. 3, 032424 (2022),
\doi{10.1103/PhysRevA.106.032424},
\url{https://link.aps.org/doi/10.1103/PhysRevA.106.032424}.

\bibitem{Marchese2023}
M. M. Marchese and P. Kok,
``Large Baseline Optical Imaging Assisted by Single Photons and Linear Quantum Optics,''
\textit{Phys. Rev. Lett.} \textbf{130}, no. 16, 160801 (2023),
\doi{10.1103/PhysRevLett.130.160801},
\url{https://link.aps.org/doi/10.1103/PhysRevLett.130.160801}.


\bibitem{vaidman1994teleportation}
Lev Vaidman.
\newblock Teleportation of quantum states.
\newblock {\em Physical Review A}, 49(2):1473, 1994.

\bibitem{braunstein1998teleportation}
Samuel~L Braunstein and H~Jeff Kimble.
\newblock Teleportation of continuous quantum variables.
\newblock {\em Physical Review Letters}, 80(4):869, 1998.

\bibitem{pirandola2006quantum}
Stefano Pirandola and Stefano Mancini.
\newblock Quantum teleportation with continuous variables: A survey.
\newblock {\em Laser Physics}, 16(10):1418--1438, 2006.

\bibitem{weedbrook2012gaussian}
Christian Weedbrook, Stefano Pirandola, Ra{\'u}l Garc{\'\i}a-Patr{\'o}n,
  Nicolas~J Cerf, Timothy~C Ralph, Jeffrey~H Shapiro, and Seth Lloyd.
\newblock Gaussian quantum information.
\newblock {\em Reviews of Modern Physics}, 84(2):621, 2012.

\bibitem{amari2023highly}
Jorge Amari, Junnosuke Takai, and Takuya Hirano.
\newblock Highly efficient measurement of optical quadrature squeezing using a
  spatial light modulator controlled by machine learning.
\newblock {\em Optics Continuum}, 2(4):933--941, 2023.

\bibitem{mandel1995optical}
Leonard Mandel and Emil Wolf.
\newblock {\em Optical coherence and quantum optics}.
\newblock Cambridge university press, 1995.

\bibitem{duan1997influence}
Lu-Ming Duan and Guang-Can Guo.
\newblock Influence of noise on the fidelity and the entanglement fidelity of
  states.
\newblock {\em Quantum and Semiclassical Optics: Journal of the European
  Optical Society Part B}, 9(6):953, 1997.

\bibitem{chizhov2001propagation}
AV~Chizhov, E~Schmidt, L~Kn{\"o}ll, and DG~Welsch.
\newblock Propagation of entangled light pulses through dispersing and
  absorbing channels.
\newblock {\em Journal of Optics B: Quantum and Semiclassical Optics}, 3(3):77,
  2001.

\bibitem{scheel2001entanglement}
Stefan Scheel, Tomas Opatrny, and D-G Welsch.
\newblock Entanglement degradation of a two-mode squeezed vacuum in absorbing
  and amplifying optical fibers.
\newblock {\em Optics and Spectroscopy}, 91(3):411--417, 2001.

\bibitem{chizhov2002continuous}
AV~Chizhov, L~Kn{\"o}ll, and D-G Welsch.
\newblock Continuous-variable quantum teleportation through lossy channels.
\newblock {\em Physical Review A}, 65(2):022310, 2002.

\bibitem{ban2002continuous}
 Masashi Ban, Masahide Sasaki,  and Masahiro Takeoka.
\newblock Continuous variable teleportation as a generalized thermalizing quantum channel
\newblock {\em Journal of Physics A: Mathematical and General}, 35(28), L401, 2002.

\bibitem{helstrom1976quantum}
CW~Helstrom.
\newblock Quantum detection and estimation theory,  academic press.
\newblock {\em New York}, 1976.

\bibitem{zhang2025}
 Yujie Zhang and Thomas Jennewein
\newblock Criteria for optimal entanglement-assisted long baseline imaging protocols.
\newblock {\em arXiv preprint arXiv:2501.16670}, 2025.



\bibitem{monras2013phase}
Alex Monras.
\newblock Phase space formalism for quantum estimation of gaussian states.
\newblock {\em arXiv preprint arXiv:1303.3682}, 2013.

\bibitem{gao2014bounds}
Yang Gao and Hwang Lee.
\newblock Bounds on quantum multiple-parameter estimation with gaussian state.
\newblock {\em The European Physical Journal D}, 68(11):1--7, 2014.

\bibitem{Huang2024}
Z. Huang, B. Q. Baragiola, N. C. Menicucci, and M. M. Wilde,
``Limited quantum advantage for stellar interferometry via continuous-variable teleportation,''
\textit{arXiv} preprint arXiv:2311.05159, 2024.


\bibitem{braunstein1994statistical}
Samuel~L Braunstein and Carlton~M Caves.
\newblock Statistical distance and the geometry of quantum states.
\newblock {\em Physical Review Letters}, 72(22):3439, 1994.

\bibitem{paris2009quantum}
Matteo~GA Paris.
\newblock Quantum estimation for quantum technology.
\newblock {\em International Journal of Quantum Information},
  7(supp01):125--137, 2009.

\bibitem{EHT2019}
Event Horizon Telescope Collaboration.
\newblock First M87 Event Horizon
Telescope results. I. The shadow of the supermassive black
hole
\newblock {\em Astrophys. J. Lett.},   875(1), L1, 2019.

\bibitem{boyajian2015stellar}
Boyajian,~Tabetha  et al.
\newblock Stellar diameters and temperatures--VI. High angular resolution measurements of the transiting exoplanet host stars HD 189733 and HD 209458 and implications for models of cool dwarfs
\newblock {\em Monthly Notices of the Royal Astronomical Society}, 447(1):846-857, 2015.

\bibitem{luo2019quantum}
Yi-Han Luo,  et al.
\newblock Quantum teleportation in high dimensions
\newblock {\em Phys. Rev, Lett.},   123(7), 070505, 2019.



\bibitem{ralph2009nondeterministic}
TC~Ralph and AP~Lund.
\newblock Nondeterministic noiseless linear amplification of quantum systems.
\newblock In {\em AIP Conference Proceedings}, volume 1110, pages 155--160.
  American Institute of Physics, 2009.

\bibitem{ralph2011quantum}
TC~Ralph.
\newblock Quantum error correction of continuous-variable states against
  gaussian noise.
\newblock {\em Physical Review A}, 84(2):022339, 2011.

\bibitem{dias2017quantum}
Josephine Dias and Timothy~C Ralph.
\newblock Quantum repeaters using continuous-variable teleportation.
\newblock {\em Physical Review A}, 95(2):022312, 2017.

\bibitem{furrer2018repeaters}
Fabian Furrer and William~J Munro.
\newblock Repeaters for continuous-variable quantum communication.
\newblock {\em Physical Review A}, 98(3):032335, 2018.

\bibitem{liu2018continuous}
Yanhong Liu, Jieli Yan, Lixia Ma, Zhihui Yan, and Xiaojun Jia.
\newblock Continuous-variable entanglement distillation between remote quantum
  nodes.
\newblock {\em Physical Review A}, 98(5):052308, 2018.

\bibitem{dias2020quantum}
Josephine Dias, Matthew~S Winnel, Nedasadat Hosseinidehaj, and Timothy~C Ralph.
\newblock Quantum repeater for continuous-variable entanglement distribution.
\newblock {\em Physical Review A}, 102(5):052425, 2020.

\bibitem{seshadreesan2020continuous}
Kaushik~P Seshadreesan, Hari Krovi, and Saikat Guha.
\newblock Continuous-variable quantum repeater based on quantum scissors and
  mode multiplexing.
\newblock {\em Physical Review Research}, 2(1):013310, 2020.

\bibitem{foreman2007remote}
Seth~M Foreman, Kevin~W Holman, Darren~D Hudson, David~J Jones, and Jun Ye.
\newblock Remote transfer of ultrastable frequency references via fiber
  networks.
\newblock {\em Review of Scientific Instruments}, 78(2):021101, 2007.

\bibitem{andersen201630}
Ulrik~L Andersen, Tobias Gehring, Christoph Marquardt, and Gerd Leuchs.
\newblock 30 years of squeezed light generation.
\newblock {\em Physica Scripta}, 91(5):053001, 2016.

\bibitem{vahlbruch2016detection}
Henning Vahlbruch, Moritz Mehmet, Karsten Danzmann, and Roman Schnabel.
\newblock Detection of 15 db squeezed states of light and their application for
  the absolute calibration of photoelectric quantum efficiency.
\newblock {\em Physical review letters}, 117(11):110801, 2016.

\bibitem{dias2022distributing}
Josephine Dias, Matthew~S Winnel, William~J Munro, TC~Ralph, and Kae Nemoto.
\newblock Distributing entanglement in first-generation discrete-and
  continuous-variable quantum repeaters.
\newblock {\em Physical Review A}, 106(5):052604, 2022.

\bibitem{brown1956correlation}
R~Hanbury Brown and Richard~Q Twiss.
\newblock Correlation between photons in two coherent beams of light.
\newblock {\em Nature}, 177(4497):27--29, 1956.

\bibitem{hale2000berkeley}
David~DS Hale, M~Bester, WC~Danchi, W~Fitelson, S~Hoss, EA~Lipman, JD~Monnier,
  PG~Tuthill, and CH~Townes.
\newblock The berkeley infrared spatial interferometer: a heterodyne stellar
  interferometer for the mid-infrared.
\newblock {\em The Astrophysical Journal}, 537(2):998, 2000.

\bibitem{caves1981quantum}
Carlton~M Caves.
\newblock Quantum-mechanical noise in an interferometer.
\newblock {\em Physical Review D}, 23(8):1693, 1981.

\bibitem{ozdemir2001quantum}
{\c{S}}ahin~Kaya {\"O}zdemir, Adam Miranowicz, Masato Koashi, and Nobuyuki
  Imoto.
\newblock Quantum-scissors device for optical state truncation: A proposal for
  practical realization.
\newblock {\em Physical Review A}, 64(6):063818, 2001.

\bibitem{pegg1998optical}
David~T Pegg, Lee~S Phillips, and Stephen~M Barnett.
\newblock Optical state truncation by projection synthesis.
\newblock {\em Physical review letters}, 81(8):1604, 1998.

\bibitem{lund2009continuous}
AP~Lund and TC~Ralph.
\newblock Continuous-variable entanglement distillation over a general lossy
  channel.
\newblock {\em Physical Review A}, 80(3):032309, 2009.

\bibitem{fiuravsek2010distillation}
Jarom{\'\i}r Fiur{\'a}{\v{s}}ek.
\newblock Distillation and purification of symmetric entangled gaussian states.
\newblock {\em Physical Review A}, 82(4):042331, 2010.

\bibitem{browne2003driving}
Daniel~E Browne, Jens Eisert, Stefan Scheel, and Martin~B Plenio.
\newblock Driving non-gaussian to gaussian states with linear optics.
\newblock {\em Physical Review A}, 67(6):062320, 2003.

\bibitem{eisert2004distillation}
J~Eisert, DE~Browne, S~Scheel, and MB~Plenio.
\newblock Distillation of continuous-variable entanglement with optical means.
\newblock {\em Annals of Physics}, 311(2):431--458, 2004.

\bibitem{hoelscher2011optimal}
Jason Hoelscher-Obermaier and Peter van Loock.
\newblock Optimal gaussian entanglement swapping.
\newblock {\em Physical Review A}, 83(1):012319, 2011.

\bibitem{fukui2021all}
Kosuke Fukui, Rafael~N Alexander, and Peter van Loock.
\newblock All-optical long-distance quantum communication with
  gottesman-kitaev-preskill qubits.
\newblock {\em Physical Review Research}, 3(3):033118, 2021.




\bibitem{pearce2017optimal}
Mark~E Pearce, Earl~T Campbell, and Pieter Kok.
\newblock Optimal quantum metrology of distant black bodies.
\newblock {\em Quantum}, 1:21, 2017.

\bibitem{yuen1983noise}
Horace~P Yuen and Vincent~WS Chan.
\newblock Noise in homodyne and heterodyne detection.
\newblock {\em Optics letters}, 8(3):177--179, 1983.

\bibitem{li1999sub}
Yong-qing Li, Dorel Guzun, and Min Xiao.
\newblock Sub-shot-noise-limited optical heterodyne detection using an
  amplitude-squeezed local oscillator.
\newblock {\em Physical review letters}, 82(26):5225, 1999.

\bibitem{li2000li}
Yongqing Li, Dorel Guzun, and Min Xiao.
\newblock Li, guzun, and xiao reply.
\newblock {\em Physical review letters}, 85(3):678, 2000.

\bibitem{ralph2000can}
TC~Ralph.
\newblock Can signal-to-noise be improved by heterodyne detection using an
  amplitude squeezed local oscillator?
\newblock {\em Physical review letters}, 85(3):677, 2000.

\bibitem{Nathan2024}
TC~Ralph.
\newblock Can signal-to-noise be improved by heterodyne detection using an
  amplitude squeezed local oscillator?
\newblock {\em Physical review letters}, 85(3):677, 2000.

@inproceedings{Nathan2024,
author = {Nathan T. Arnold and Colin P. Lualdi and Michael E. Goggin and Paul G. Kwiat},
title = {{All-optical quantum memory}},
volume = {12911},
booktitle = {Quantum Computing, Communication, and Simulation IV},
editor = {Philip R. Hemmer and Alan L. Migdall},
organization = {International Society for Optics and Photonics},
publisher = {SPIE},
pages = {129111C},
keywords = {Quantum memory, Photonic memory, Delay line, Herriott cell, Robert cell},
year = {2024},
doi = {10.1117/12.3003228},
URL = {https://doi.org/10.1117/12.3003228}


Arnold, N. T., Lualdi, C. P., Goggin, M. E., & Kwiat, P. G. (2024, March). All-optical quantum memory. In Quantum Computing, Communication, and Simulation IV (Vol. 12911, pp. 390-394). SPIE.


\end{thebibliography}

\end{document}